\documentclass[twocolumn]{aastex62}
\usepackage{amsmath}
\usepackage{lineno}

\newcommand{\Msun}{\ensuremath{{\rm M}_\odot}}
\newcommand{\mstar}{\ensuremath{M_*}}
\newcommand{\rstar}{\ensuremath{R_*}}
\newcommand{\Rsun}{\ensuremath{{\rm R}_{\odot}}}

\newcommand{\mdotbondi}{\ensuremath{\dot{M}_{\rm B}}}
\newcommand{\rbondi}{\ensuremath{R_{\rm B}}}
\newcommand{\rhoagn}{\rho_{\rm AGN}}
\newcommand{\csa}{c_{s,{\rm AGN}}}

\newcommand{\agns}{AGN stars}

\graphicspath{{./}{figures/}}

\submitjournal{ApJ}

\shorttitle{Compact Objects within AGN Stars}
\shortauthors{Cantiello et al.}

\begin{document}
\title{Signatures of Compact Object Mergers Inside Stars in AGN Disks}

\correspondingauthor{Matteo Cantiello}
\email{mcantiello@flatironinstitute.org}

\author[0000-0002-8171-8596]{Matteo Cantiello}
\affil{Center for Computational Astrophysics, Flatiron Institute, 162 5th Avenue, New York, NY 10010, USA}
\affil{Department of Astrophysical Sciences, Princeton University, Princeton, NJ 08544, USA}
\author[0000-0001-6157-6722]{Alexander J. Dittmann}
\altaffiliation{NASA Einstein Fellow}
\affiliation{Institute for Advanced Study, 1 Einstein Drive, Princeton, NJ 08540, USA}
\author[0000-0002-5956-851X]{Saavik Ford}
\affil{Center for Computational Astrophysics, Flatiron Institute, 162 5th Avenue, New York, NY 10010, USA}
\affiliation{Department of Science, BMCC, City University of New York, New York, NY 10007, USA}
\author[0000-0002-9726-0508]{Barry McKernan}
\affil{Center for Computational Astrophysics, Flatiron Institute, 162 5th Avenue, New York, NY 10010, USA}
\affiliation{Department of Science, BMCC, City University of New York, New York, NY 10007, USA}
\author[0000-0003-0634-531X]{Carlos Palenzuela}
\affil{Departament de F\'isica, Universitat de les Illes Balears, Palma de Mallorca, E-07122, Spain}
\affil{Institute of Applied Computing \& Community Code (IAC3),
Universitat de les Illes Balears, Palma de Mallorca, E-07122, Spain}
\author[0000-0002-3635-5677]{Rosalba Perna}
\affil{Department of Physics and Astronomy, Stony Brook University, Stony Brook, NY 11794-3800, US}
\author[0000-0003-2012-5217]{Taeho Ryu}
\affil{JILA, University of Colorado and National Institute of Standards and Technology, 440 UCB, Boulder, CO 80309-0440, USA}
\affil{Department of Astrophysical and Planetary Sciences, 391 UCB, University of Colorado, Boulder, CO 80309-0391, USA}

\begin{abstract}
Disks of gas accreting onto supermassive black holes, powering active galactic nuclei (AGN), can capture stars from nuclear star clusters or form stars in situ via gravitational instability. The dense, hot disk environment can drive rapid accretion onto embedded stars, dramatically altering their evolution. Models predict that, for sufficiently rapid accretion, fresh gas replenishes hydrogen in stellar cores as quickly as it is burned, and the stars reach a quasi-steady state. Here we study encounters of such massive, long-lived (``immortal'') stars with compact objects in AGN disks. We estimate the encounter rate and the timescale for a single compact remnant to spiral into an AGN star; depending on how strongly feedback regulates the ensuing accretion, the star is either consumed in a collapsar-like, engine-driven transient or converted into a long-lived, quasi-star-like object hosting a central black hole. We then study the merger of a binary black hole (BBH) inside the AGN star, and show that gas drag hardens the binary to merger far faster than gravitational-wave emission alone. The resulting merger is a loud LIGO--Virgo--KAGRA (LVK) source, but the characteristic imprint of the dense environment---a strong suppression and dephasing of the inspiral relative to vacuum---falls in the deci-Hz band rather than the milli-Hz LISA band, and is best resolved by next-generation detectors such as DECIGO. We estimate that this channel could contribute a BBH merger rate of up to $\sim8\,{\rm Gpc^{-3}\,yr^{-1}}$ in favorable scenarios, and discuss the model uncertainties and directions for future work.
\end{abstract}


\section{Introduction} \label{sec:intro}
Active galactic nuclei (AGN) are powered by the accretion of massive gas disks onto supermassive black holes \citep{1969Natur.223..690L,2008ARA&A..46..475H}. The outer regions of these disks can become gravitationally unstable, leading to star formation \citep[e.g.][]{1980SvAL....6..357K,2003MNRAS.339..937G,2020MNRAS.493.3732D,2022arXiv220510382D}. 
Furthermore, stars can be captured by the disk from nuclear star clusters, which are ubiquitous at least in quiescent galaxies \citep[e.g.][]{Neumayer20}, through a variety of mechanisms.
Gas torques may capture initially misaligned stars into the disk \citep[e.g.][]{1993ApJ...409..592A,1995MNRAS.275..628R}, further align stellar orbits to the midplane of the disk \citep[e.g.][]{2004ApJ...602..388T}, and circularize initially eccentric stellar orbits \citep[e.g.][]{1991MNRAS.250..505S,2020ApJ...889...94M}.

Regardless of the mechanisms by which stars become embedded in AGN disks, accretion from the disk can proceed at a high rate \citep{Chen:2024,Chen:2025} and profoundly alter their evolution \citep{Cantiello:2021,2021ApJ...916...48D,2021ApJ...914..105J,Dittmann:2025,Dittmann2026}, leading to the formation of very massive stars \citep[e.g.][]{Cantiello:2021,2021ApJ...911L..14W}, gamma-ray bursts \citep[e.g.][]{Perna:2021,2021ApJ...914..105J,2021ApJ...911L..19Z,2021MNRAS.508.1842D,Lazzati:2023}, and gravitational wave event progenitors \citep[e.g.][]{1999ApJ...521..502C,Stone17,2022ApJ...929..133J,2022arXiv220510382D}. Analytical and evolutionary models which embed stars in a medium of constant density and temperature predict that stars may reach a quasi-steady state, undergoing no chemical evolution \citep{Cantiello:2021,2021ApJ...916...48D,2023ApJ...946...56D,Dittmann:2025,Dittmann2026}. Stars reach such a state (``immortal'') when the accretion of fresh gas can supply hydrogen to stellar cores faster than it is spent powering the star, leading to a balance between accretion and radiation-driven mass loss. 

Immortal stars may, in principle, be able to survive as long as the AGN disk persists. Observations indicate that individual AGN accretion episodes can be as short as $\sim10^{5}$ years \citep{2015MNRAS.451.2517S,10.1093/mnrasl/slv098}, although galactic nuclei are expected to be active for an integrated duration $\mathcal{O}(10^{8})$ years over a Hubble time \citep[e.g.][]{2004cbhg.symp..169M}. The median AGN episode lifetime is still unknown, although in the local Universe it is likely to be $<10\,{\rm Myr}$ \citep[e.g.][]{Oppenheimer18}. 

In nuclear star clusters, a large central population of stellar-mass black holes (BHs), with number density $\sim10^{4}\,{\rm pc}^{-3}$, is expected \citep[e.g.][]{Morris93, Miralda00, Generozov18}. As with stars, some fraction of nuclear BHs have orbits that reside within the AGN disk plane, and an additional fraction are captured over the disk lifetime \citep[e.g.][]{Fabj20}. Moreover, BHs can be formed in the disk by AGN stars as well \citep{Cantiello:2021}. The resulting population of BHs embedded in AGN disks can migrate, form binaries and merge, yielding a promising source of some of the binary black hole (BBH) mergers observed by LIGO--Virgo--KAGRA (LVK) in gravitational waves (GW) \citep{McK14,Bartos17,Stone17}; see \citet{FordMcK26} for a recent review. In particular, predictions of the LVK AGN channel include very efficient formation of intermediate-mass black holes (IMBHs) \citep[e.g.][]{McK12,Secunda19,Tagawa21}, asymmetric mass mergers and possibly eccentricity in the LVK band \citep{Samsing22}. 

A fraction of BH mergers observed by LVK lie in the upper-mass gap ($\sim 50$--$120M_{\odot}$), which is most easily explained by a hierarchical merger origin in a deep potential well \citep[e.g.][]{GerosaFishbach21,Ford22}. Effective spins of BHs in observed mergers are not centered on $\chi_{\rm eff} \sim 0$ as expected in a gas-free hierarchical merger scenario, but they are rather biased towards $\chi_{\rm eff} >0$ \citep{Abbott21}; population modeling of the current gravitational wave transient catalog suggests that between $\sim9$--$40\%$ of mergers must originate from a preferentially aligned channel to explain the observed asymmetry \citep{2026arXiv260527226T}. In the context of dynamical models, this suggests a symmetry-breaking agent (such as an AGN disk) \citep[e.g.][]{Wang21,McK23}.

If immortal stars exist, given their high masses they should migrate rapidly through AGN disks \citep{Bellovary16}, encountering embedded compact objects. Here we investigate the nature of BH-immortal star encounters in AGN disks and how mergers of BHs may be facilitated by immortal stars. We consider the possible imprints of the immortal channel on GW merger signatures and other observables.

\section{AGN Star Structure}\label{sec:structure}
AGN stars at rest relative to the gas are characterized by an accretion stream of material captured in the disk from the Bondi radius
\begin{equation}
\rbondi = \frac{2G\mstar}{\csa^2},
\label{eq:rbondi}
\end{equation}
where  $\csa$ is the local sound speed in the disk and $\mstar$ is the stellar mass. If the local AGN disk density is $\rhoagn$ then the AGN star accretes material at a rate
\begin{equation}
\mdotbondi =  \eta \, \pi \rbondi^2 \,\rhoagn\, \csa,
\label{eq:mbondi}
\end{equation}
where $\eta$ is an efficiency factor ($\eta \le 1$). In \agns\ the ratio $\rbondi/\rstar$ is $\sim 10^{2}$--$10^{4}$ for typical parameters \citep{Cantiello:2021}, and reaches $\sim2\times10^{5}$ for the massive immortal model adopted here.

\begin{figure*}
\includegraphics[width=\linewidth]{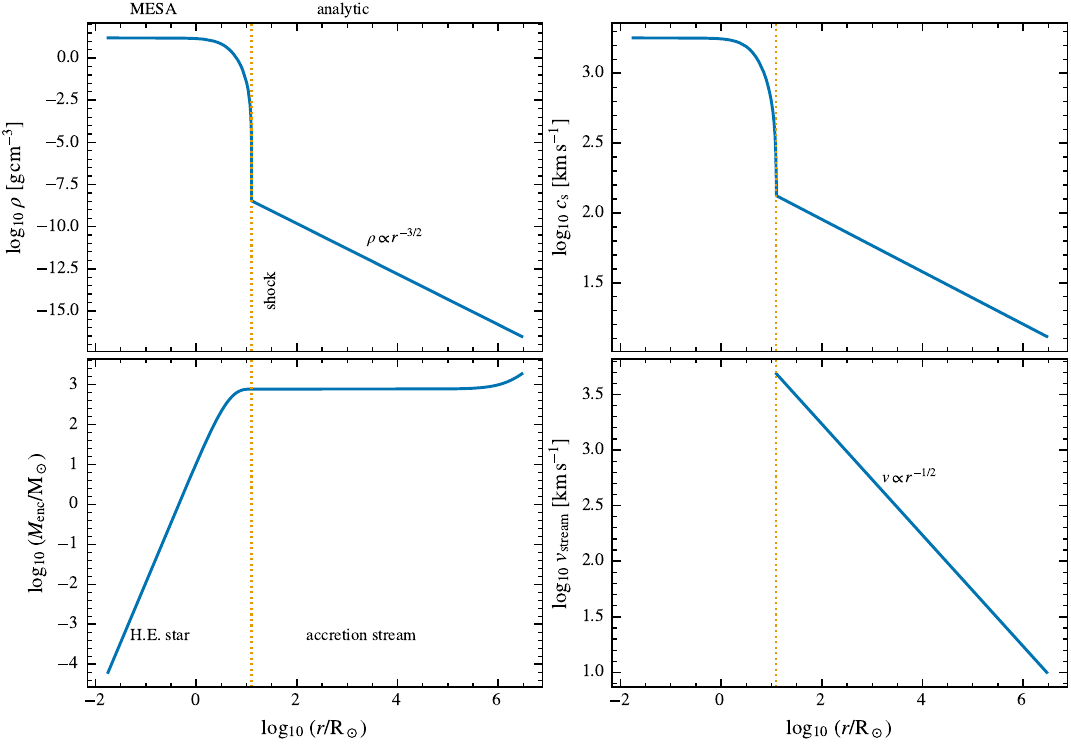}
\caption{Structure of an AGN star, from its center to the Bondi radius. The model is divided into two parts: from the stellar center to the accretion shock radius the star is in hydrostatic equilibrium and calculated using \texttt{MESA}. The properties of the accretion stream from the Bondi radius to the shock radius are calculated using analytical approximations, and provide outer boundary conditions to the \texttt{MESA} calculation \citep{Cantiello:2021}. Density and sound speed at the Bondi radius match the AGN disk conditions at the location of the AGN star.}
\label{fig:structure}
\end{figure*}

\subsection{Numerical Methods}\label{sec:numerical}
We model the evolution of AGN stars using revision r22.05.1 of the Modules for Experiments in Stellar Astrophysics \citep[\texttt{MESA};][]{{2011ApJS..192....3P},{2013ApJS..208....4P},{2015ApJS..220...15P},{2018ApJS..234...34P},{2019ApJS..243...10P},{2022arXiv220803651J}} software instrument. We implement accretion, mass loss, and modified boundary conditions following \citet{Cantiello:2021}, including, e.g., treatment of effects such as the ram pressure of the accretion stream onto the star and irradiation of the star by the AGN disk \citep[e.g.][]{Cantiello:2021,2021ApJ...914..105J,2021ApJ...916...48D}.

As in previous works, we relax the atmospheric boundary conditions and accretion rate sequentially such that the atmospheric boundary conditions are fully relaxed before accretion begins, over a period of approximately $10^7$ years \citep[e.g.][]{Cantiello:2021,2021ApJ...916...48D}. The total relaxation time is sufficiently short that it amounts to only a small fraction of the total evolutionary time of our initial $1\,M_\odot$ models. Additionally, as in previous work, we assume an enhancement of compositional mixing in radiative regions as models approach the Eddington luminosity \citep[e.g.][]{Cantiello:2021}. Our \texttt{MESA} models are non-rotating; rotation enters the discussion only through the angular momentum of the accreted gas (\S\ref{sec:orbital}, \S\ref{sec:bbh_props}), and a self-consistently rotating structure is left to future work.

Stars in AGN disks can become very massive and reach a quasi-steady state, undergoing no chemical evolution (``immortal''). This occurs because the
accretion of fresh gas can supply hydrogen to the stellar core faster than it can be burned \citep{Cantiello:2021}. This type of evolution is expected in a fairly broad region of the AGN disk \citep{Fabj:2025,Dittmann:2025,Dittmann2026}. 

\subsection{AGN star profile}

Fig.~\ref{fig:structure} shows the typical structure of an immortal AGN star, illustrating (top panels) the profiles of density and sound speed, and correspondingly (bottom panels), the enclosed mass and accretion stream velocity. The star is characterized by a central region in hydrostatic equilibrium, with a radius of about $12.5\,R_{\odot}$ enclosing a total mass $M_{\ast}\simeq773\,M_{\odot}$ and a central density $\rho_{\rm c}\simeq16\,{\rm g\,cm^{-3}}$. Material accretes (via an accretion stream) onto this central object from the Bondi radius down towards an inner point where a shock is formed. The size of the Bondi radius varies with the location of the star within the AGN disk, as it depends on the local density and sound speed. The enclosed mass $M_{\rm en}(r)$, density, pressure, and sound speed used in the dynamical calculations of \S\ref{sec:results} are read directly from the MESA profile (using the MESA mass coordinate $M(<r)$ rather than reconstructing it from $\rho(r)$). Unless stated otherwise we adopt the gas-pressure sound speed $c_{s}=\sqrt{(5/3)\,P_{\rm gas}/\rho}$ as fiducial, and consider the radiation-inclusive adiabatic value where it matters (\S\ref{sec:bbh}).

Note that the drag calculations of \S\ref{sec:results} use the specific $773\,M_{\odot}$ MESA model of Fig.~\ref{fig:structure}, whereas the Hill-sphere, rate, and migration estimates (\S\ref{sec:bbh}, \S\ref{sec:rates}) are normalized to a generic $\mathcal{O}(100\,M_{\odot})$ immortal for ease of comparison with the AGN-disk migration literature. The two choices are consistent: rescaling the migration time to the $773\,M_{\odot}$ model would shorten $\tau_{\rm mig}\propto M_{\ast}^{-1}$ by a factor $\sim7.7$, strengthening rather than weakening the encounter rate.

\section{Results}\label{sec:results}

We study the dynamical interaction of compact objects delivered to the stellar interior. First we focus on a \emph{single} compact object that sinks toward the center under gas drag (\S\ref{sec:orbital}), then we study an embedded \emph{binary} black hole (BBH) whose internal orbit hardens within the dense core toward merger (\S\ref{sec:bbh}). In \S\ref{sec:perturb} we identify where the compact object can be treated as a small perturbation to the stellar background, and where that approximation fails---most notably during the inner binary-hardening phase---and in \S\ref{sec:stream} we show that the accretion stream feeding the star does not by itself capture or appreciably decelerate compact objects.

Three configurations appear in what follows. (i) A \emph{single BH} sinking through an otherwise BH-free star (\S\ref{sec:orbital}; Figs.~\ref{fig:inspiral}, \ref{fig:spiral}a). (ii) A BH sinking onto a companion left at the stellar center by an earlier capture (\emph{sequential capture}), which we use as the concrete two-BH geometry in Figs.~\ref{fig:power} and \ref{fig:spiral}b. (iii) A pre-assembled binary entering the star and sinking as a single heavier perturber (\emph{direct capture}). Routes (ii) and (iii) deliver the same tightly bound pair to the dense core, where the internal hardening of \S\ref{sec:bbh} takes over and is insensitive to the formation route.

\subsection{Does the BH perturb the stellar structure?}
\label{sec:perturb}

Before beginning our investigation of the effects of a BH/BBH inside an AGN star, we need to evaluate the conditions under which the structure of the star can be considered not significantly altered by the presence of the BH. The two most important potential sources of disruption come from the gravitational influence of the BH, and from the release of energy due to accretion.

The ratio of the Bondi-Hoyle-Lyttleton (BHL) radius $r_{\rm BHL}=2GM_{\rm BH}/(v^{2}+c_{s}^{2})$ to the orbital radius $r$, evaluated for the BH's supersonic circular motion $v^{2}\simeq GM_{\rm en}(r)/r\gg c_{s}^{2}$, reduces to a pure mass ratio in which the dimensional factors cancel,
\begin{align}
    \frac{r_{\rm BHL}}{r} \simeq \frac{2M_{\rm BH}}{M_{\rm en}(r)} = 4\times 10^{-2}\left(\frac{M_{\rm BH}}{10M_{\odot}}\right)\left(\frac{M_{\rm en}(r)}{500M_{\odot}}\right)^{-1}.
\end{align}
Thus, the BH cross-section is very small compared to the size of the star. The two radii become comparable only near the inner core ($r\sim R_{\odot}$, where $M_{\rm en}(r)\sim2M_{\rm BH}$). The distance at which the BH tidally disrupts the central part of the core is \citep{1975Natur.254..295H,Ryu+2020}
\begin{align}
    r_{\rm t,c} \simeq \left(\frac{M_{\rm BH}}{\rho_{\rm c}}\right)^{1/3} \simeq 1.5\, R_{\odot}\left(\frac{M_{\rm BH}}{10M_{\odot}}\right)^{1/3}\left(\frac{\rho_{\rm c}}{16~{\rm g ~cm}^{-3}}\right)^{-1/3};
\end{align}
Note that $r_{\rm t,c}$ is, up to a factor, the radius enclosing $\sim M_{\rm BH}$ of core gas, so it roughly coincides with the mass-ratio breakdown above. Inside this radius the BH shreds and consumes the remaining core material, seeding the central hyper-Eddington accretion discussed in \S\ref{sec:orbital}. Because the orbital decay is slow compared with the stellar dynamical time (their ratio is $\mathcal{O}(100)$), the star has time to relax back toward equilibrium as the BH sinks, so its structure is not strongly perturbed during the single-object envelope inspiral. This no longer fully holds when the in-spiraling object is a BBH rather than a single BH: during the inner binary-hardening phase the deposited energy does become comparable to the envelope binding energy (\S\ref{sec:bbh}).

Because these BHs move through optically thick gas media, photons generated by accretion may be trapped and advected onto the BH rather than escaping \citep{Begelman1979}. Such photons must diffuse out across the Bondi sphere, of size $r_{\rm BHL}$, on the diffusion time $t_{\rm diff}\simeq r_{\rm BHL}\tau/c$, where $\tau\simeq \rho_{\rm B}r_{\rm BHL} \kappa$ is the optical depth and $\rho_{\rm B}$ the density at the Bondi radius. They are advected with the flow if $t_{\rm diff}$ exceeds the accretion time---the time for the gas to fall from the Bondi radius onto the BH, $t_{\rm acc}\simeq\sqrt{r_{\rm BHL}^{3}/G M_{\rm BH}}$. Using $r_{\rm BHL}=2GM_{\rm BH}/v^{2}\simeq 2(M_{\rm BH}/M_{\rm en})r$ for the supersonic orbital velocity $v^{2}\simeq G M_{\rm en}/r$, the ratio is
\begin{align}
    \frac{t_{\rm diff}}{t_{\rm acc}}&
    \simeq
    6\times10^{2} \left(\frac{M_{\rm BH}}{10\Msun{}}\right)\left(\frac{\kappa}{0.34~{\rm cm}^{2}~{\rm g}^{-1}}\right)\nonumber\\
    &\times
    \left(\frac{\rho_{\rm B}}{10^{-5}~{\rm g ~cm}^{-3}}\right)\left(\frac{r}{8\times10^{11}{\rm ~cm}}\right)^{1/2}\left(\frac{M_{\rm en}}{773M_{\odot}}\right)^{-1/2},
\end{align}
where the fiducial values correspond to conditions just beneath the stellar surface---the point along the inspiral where the trapping is weakest. The ratio greatly exceeds unity everywhere and is a conservative estimate, since $\tau$ is evaluated with the lower density at the outer edge of the Bondi sphere, so the photons are swept onto the BH before they can escape: the accretion is radiatively inefficient, its energy advected inward rather than radiated. The trapping strengthens toward smaller radii, where the density rises steeply (Fig.~\ref{fig:structure}): evaluated along the model profile, $t_{\rm diff}/t_{\rm acc}$ grows from $\sim5\times10^{2}$ near the surface to $\gtrsim10^{8}$ in the dense interior.

\subsection{The accretion stream}
\label{sec:stream}

Figure~\ref{fig:structure} also shows the accretion stream that feeds the immortal, extending from the accretion shock at the stellar surface ($R_{\ast}\simeq12.5\,\Rsun$) out to the Bondi radius $R_{\rm B}\simeq3\times10^{6}\,\Rsun\simeq0.07$~pc, where the inflow joins the ambient AGN disk. Strictly, at the fiducial disk locations of \S\ref{sec:bbh} the star's Hill radius and the disk scale height are both far smaller than $\rbondi$, so the SMBH tide and the disk's vertical structure truncate the star-dominated inflow well inside $\rbondi$. The efficiency factor $\eta$ in Eq.~(\ref{eq:mbondi}) absorbs this truncation, and the analytic stream of Fig.~\ref{fig:structure} should be read as applying inside $\min(R_{\rm Hill},H)$. Although this stream supplies the star, our estimates indicate that it is dynamically subdominant for compact-object capture and inspiral, so that the immortal's influence on a captured BH begins effectively at the stellar surface.

The stream is far too tenuous to capture or decelerate a BH. Its density falls steeply outward  as $\rho\propto r^{-3/2}$ (for the specific AGN star model adopted here it ranges from $3\times10^{-9}\,{\rm g\,cm^{-3}}$ just beyond the shock to the disk value $3\times10^{-17}\,{\rm g\,cm^{-3}}$ at $R_{\rm B}$), so the gas enclosed within the BH's own Bondi radius $R_{\rm B,BH}=2Gm_{\rm BH}/(v^{2}+c_{s}^{2})$ is only $\lesssim10^{-5}$ of the BH mass throughout the stream. Integrating the gas drag (dynamical friction plus accretion) along a radial passage through the stream, a $10\,\Msun$ BH loses only $\lesssim10^{-5}$ of its kinetic energy per passage. The geometry most favorable to capture is a near-co-orbital encounter, which arrives with a relative energy of only $\tfrac{1}{2}v_{\rm shear}^{2}$ and therefore requires a fractional energy loss of just $\sim(v_{\rm shear}/v_{\rm peri})^{2}\sim10^{-3}$--$10^{-4}$ at pericenter. Even then, of order $10^{2}$--$10^{3}$ stream passages would be needed, so the stream contributes at most marginally to the capture rate. These estimates conservatively use the BH's own velocity: the stream is itself in free-fall ($v_{\rm str}\simeq\sqrt{2GM/r}\simeq\sqrt{2}\,v_{\rm kep}$), and since the gravitational drag scales as $\rho/(v_{\rm rel}^{2}+c_{s}^{2})$, including the relative motion only weakens it further---by a factor of a few for an orbiting BH, and to zero for one co-infalling with the stream. The localized stream mass is similarly negligible: within $\sim10^{3}\,R_{\ast}$ of the star it amounts to only $\sim0.3\,\Msun\ll m_{\rm BH}$ (and reaches just $\sim10\,\Msun$ by $\sim10^{4}\,R_{\ast}$), while the much larger spherically-integrated value out to $R_{\rm B}$ ($\sim10^{3}\,\Msun$) is merely ambient disk gas filling the enormous Bondi sphere rather than a localized perturber. We therefore attribute the capture of a BH into the stellar vicinity to the broader AGN-disk dynamics via migration and gas-assisted encounters \citep[e.g.,][]{Fabj20}, and adopt the stellar surface as the starting point of the inspiral computed below.

\subsection{AGN Star + BH: Orbital Evolution}
\label{sec:orbital}

A compact object that becomes bound to an immortal sinks through the stellar envelope on a quasi-circular orbit, transferring its orbital energy and angular momentum to the gas. This sinking of a compact object through a stellar envelope is the AGN-disk analog of a common-envelope inspiral, whose gravitational-wave (GW) signature and prospects for detection with LISA were studied by \citet{Renzo:2021}. Three dissipative channels act on the inspiraling BH: gaseous dynamical friction (DF) against the stellar material, the drag associated with BHL accretion onto the BH, and gravitational-wave emission. We compute the DF force following the circular-orbit prescriptions of \citet{Ostriker99} and \citet{Kim207}, the accretion drag as $\dot{M}_{\rm BHL}\,v$ with the BHL rate (momentum conservation of the accreted gas), and the GW luminosity from \citet{Peters64}. The accretion-drag term is an upper bound: the accreted gas carries angular momentum and should circularize into a disk around the BH, and the associated feedback (see below) will reduce the effective capture rate. We also neglect the mass the BH would gain at face-value BHL rates. Dynamical friction dominates the drag budget everywhere (Fig.~\ref{fig:power}), so neither assumption appreciably affects the sink timescale. A gas-captured BH also generally enters on an eccentric rather than circular orbit; pericenter drag in the denser inner layers then removes energy even faster, so the quasi-circular assumption is conservative.

Because the BH orbits the mass enclosed within its orbit, $M_{\rm en}(r)$, rather than a central point mass, we follow the decay through the orbital angular momentum $J = m_{\rm BH}\sqrt{G M_{\rm en}(r)\,r}$. A tangential dissipative channel of luminosity $L_{i}$ removes angular momentum at the rate $L_{i}/\Omega$ (power), where $\Omega(r)=\sqrt{G M_{\rm en}(r)/r^{3}}$ is the local orbital frequency, so the equation of motion for the sinking BH is
\begin{equation}
    \frac{dJ}{dt} = -\frac{L_{\rm DF}+L_{\rm acc}+L_{\rm GW}}{\Omega(r)},
    \label{eq:dLdt}
\end{equation}
which we integrate inward to obtain the trajectory $r(t)$. The gas-drag terms dominate the GW term by many orders of magnitude throughout most of the stellar interior, so the decay is effectively gas-drag-driven.

\begin{figure}
\centering
\includegraphics[width=\columnwidth]{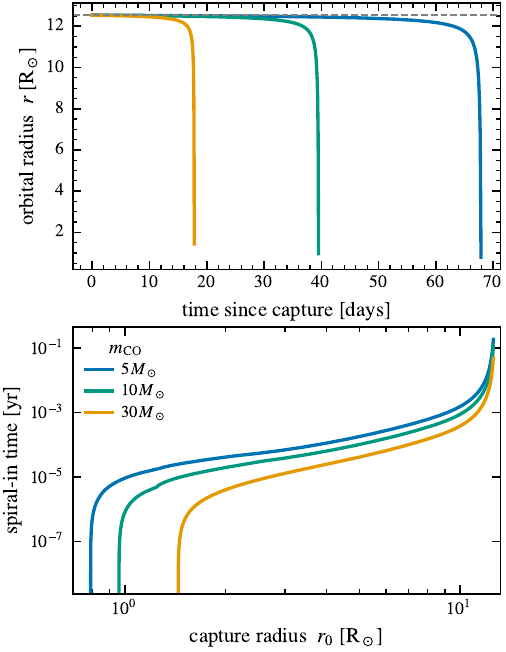}
\caption{Spiral-in of a single compact object captured by an immortal, obtained by integrating Eq.~(\ref{eq:dLdt}). \emph{Top:} orbital radius $r(t)$ for $5$, $10$, and $30\Msun$ black holes released near the stellar surface; all reach the core within tens of days, with more massive objects sinking faster. \emph{Bottom:} spiral-in time as a function of capture radius $r_{0}$. The timescale is set by the tenuous outer envelope, so a BH captured deeper in the star sinks almost instantaneously.}
\label{fig:inspiral}
\end{figure}

\begin{figure*}
\centering
\includegraphics[width=\linewidth]{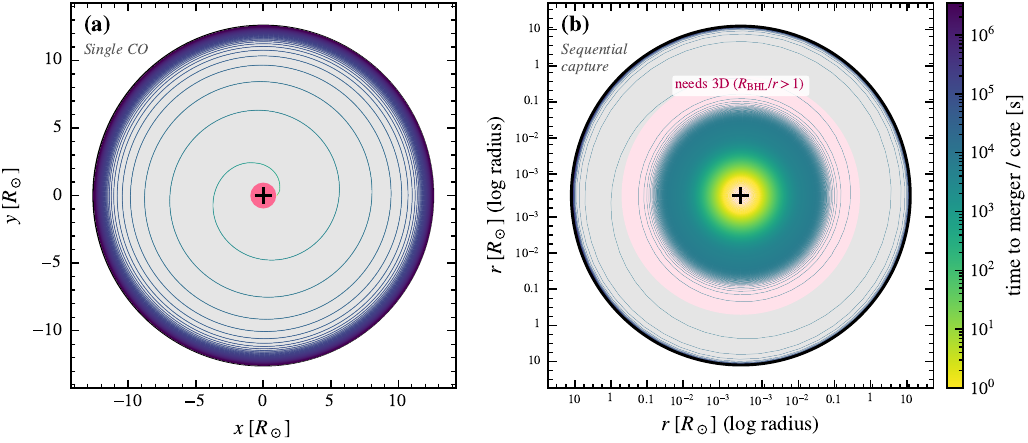}
\caption{Spiral-in trajectories from integrating Eq.~(\ref{eq:dLdt}), plotted as
$(r\cos\phi,\,r\sin\phi)$ with the accumulated orbital phase
$\phi(t)=\int\Omega\,dt$; color gives the (log-scaled) time remaining to the
endpoint, from $\sim\!0.1$~yr near the surface to seconds at the end.
{\bf (a)} A single $10\,\Msun$ BH captured at the stellar surface
($R_{\ast}\simeq12.5\,\Rsun$), on a linear radial scale: it lingers in the
tenuous outer envelope for most of its few-hundred orbits, then plunges through
the dense interior. The pink disk marks $r\lesssim0.96\,\Rsun$, where the
enclosed stellar mass drops below the BH mass and the test-particle treatment
fails.
{\bf (b)} The sequential-capture configuration (\S\ref{sec:bbh}): the same
inspiral with an equal-mass BH held at the center, followed on a logarithmic
radial scale to merger. The drag physics is identical to (a), with the GW
luminosity of the BH pair added. The orbit winds several thousand times,
crowding near the gas--GW crossover at a few~$\times10^{-3}\,\Rsun$---where the
measurable deci-Hz dephasing accumulates (\S\ref{sec:gwobs})---before plunging
to the ISCO. Here the pink disk marks the region $R_{\rm BHL}>r$
($r\lesssim0.51\,\Rsun$), where the point-perturber treatment fails and a 3D
self-gravitating calculation is required. This panel is illustrative; the
quantitative hardening times and waveforms use the fiducial model of
\S\ref{sec:bbh} and Appendix~\ref{apndx:ladder}.}
\label{fig:spiral}
\end{figure*}

\begin{figure*}
    \centering
    \includegraphics[width=\linewidth]{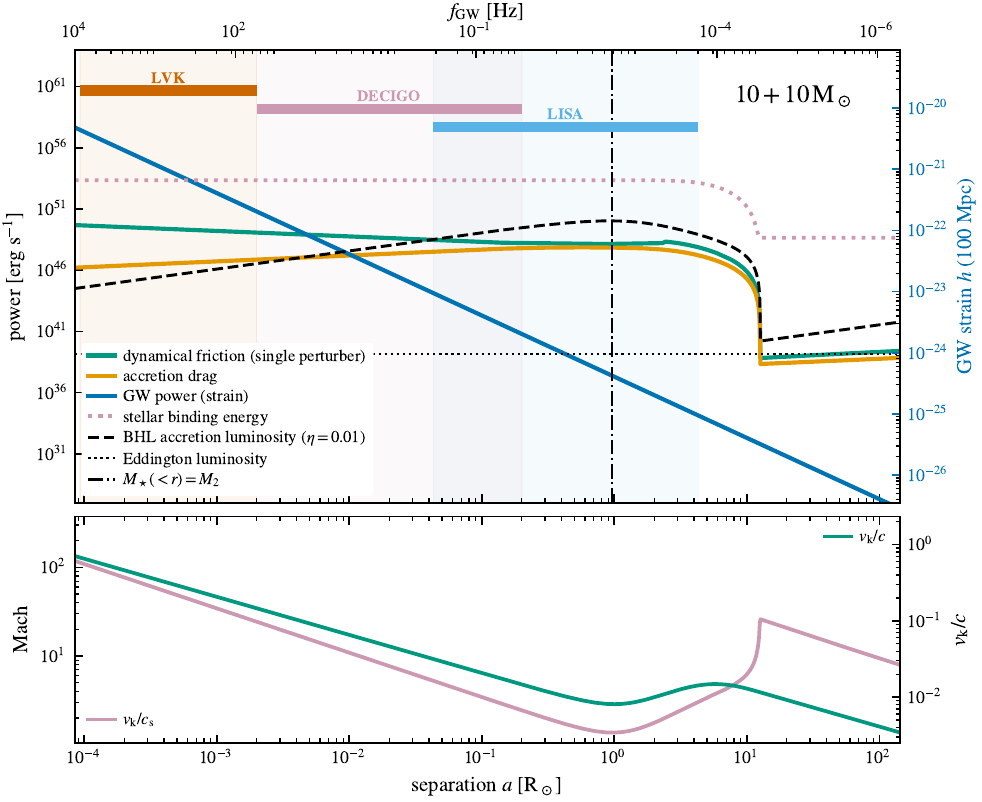}
    \caption{Power extracted by each dissipative channel for a BBH spiraling inside the AGN star, comparing the gas-drag and gravitational-wave channels on the same footing (the single compact object of Fig.~\ref{fig:inspiral} instead sinks through an otherwise BH-free interior and emits no comparable GW power). Following an earlier capture, a BH of mass 10$\Msun$ (BH1) sits at the center of the AGN star (stellar mass 773$\Msun$); a second, equal-mass BH (BH2) is captured and spirals inwards. The upper plot shows, as a function of separation from the center of the star and BH1, the single-perturber dynamical-friction drag power on BH2 from its interaction with the stellar gas, the accretion drag power from momentum conservation of material accreting onto BH2 at the BHL rate, and the GW  power generated by the inspiral of BH2 around BH1. The top axis gives the corresponding GW frequency, with the nominal LISA, DECIGO, and LVK bands shaded; the right axis re-expresses the GW power curve as characteristic strain at $d = 100$~Mpc (using $h \propto L_{\rm GW}^{1/5}$). We also show the BHL accretion luminosity assuming a conversion efficiency $\eta = 0.01$; this rate likely over-estimates the true accretion onto BH2, as suggested by the horizontal dotted line marking the Eddington luminosity of BH2. The binding energy of the AGN star (which does not include any black hole) is shown by the purple dotted line. The vertical line marks the radial coordinate enclosing a stellar mass equal to 10$\Msun$. The bottom plot shows the ratio of the orbital velocity $v_{\rm k}(a)$ (defined in \S\ref{sec:orbital}) to the local gas-only sound speed of the unperturbed stellar material (Mach number), and to the speed of light; the GW frequency on the top axis uses the binary orbital frequency, which coincides with $v_{\rm k}$ in the deep core where the detectable GW emission arises.}\label{fig:power}
\end{figure*}

Figure~\ref{fig:inspiral} shows the resulting spiral-in. A $10\Msun$ BH captured near the stellar surface ($R_{\ast}\simeq 12.5\Rsun$) reaches the core in $\sim 0.1\,{\rm yr}$, with heavier objects sinking faster (the drag scales steeply with mass). The inspiral time is dominated by the tenuous outer envelope, where the density, and hence the drag, is lowest: the BH lingers in the outer layers and then rapidly plunges through the dense interior, so the capture radius largely sets the timescale. Linear-response calculations in stratified envelopes \citep{Gagnier26} indicate that the global wake enhances the azimuthal drag over these local, uniform-medium prescriptions by a factor $\sim1$--$3$ across the envelope (a result we confirmed for our specific stellar model with that work's public code), so the spiral-in times quoted here are conservative upper estimates. Figure~\ref{fig:spiral}(a) shows this geometry: the BH spirals from the surface down to a small central core (shaded pink, $r\lesssim0.96\,\Rsun$), where the enclosed stellar mass drops below the BH mass and our test-particle treatment fails. The companion Fig.~\ref{fig:spiral}(b) panel follows the same configuration with a second black hole held at the center, on a logarithmic scale all the way to merger; it is discussed in \S\ref{sec:bbh}.

The orbital energy released during the sink, $\sim G m_{\rm BH} M_{\rm en}(r)/2r \sim 2\times10^{50}\,{\rm erg}$ once $M_{\rm en}(r)\sim m_{\rm BH}$, is only $\sim 0.1\%$ of the stellar gravitational binding energy ($\sim2\times10^{53}\,{\rm erg}$); drag heating alone therefore cannot unbind the immortal. Accretion can, however, provided we distinguish between the rest-mass energy that is \emph{liberated} and the fraction that \emph{escapes} into the envelope. At the central density the BHL rate is enormously super-Eddington, and accreting only $\sim 1\Msun$ liberates the full binding energy for a nominal conversion efficiency $\eta\simeq0.1$ (considerably less mass suffices once the internal energy of the radiation-dominated envelope is accounted for; \S\ref{sec:noimmortal}). Only the escaping fraction can unbind stellar material, though, and the two efficiencies diverge strongly in this regime. The flow is photon-trapped (\S\ref{sec:perturb}), and radiation-GRMHD simulations of super-Eddington accretion confirm that the escaping radiative efficiency falls well below the nominal $\eta$ as the accretion rate rises, with much of the dissipated energy advected across the horizon \citep{Zhang25a,Zhang25b}. Energy can still be extracted mechanically once the inflow is mediated by an accretion disk, whose jets and disk winds carry energy as bulk motion and magnetic stress and are therefore not subject to photon trapping. This mechanical efficiency depends on the BH spin and the accumulated magnetic flux, and it is the dominant uncertainty in the destruction energetics. Such a disk is expected: stars in AGN disks accrete high-angular-momentum gas and rotate rapidly \citep{2021ApJ...914..105J}, so the material captured by the sunk BH carries enough angular momentum to circularize into a centrifugally supported disk rather than accrete radially. The fate of a star harboring a single sunk BH therefore depends on how effectively this mechanical feedback regulates the inflow, and is bracketed by two regimes. If regulation fails and accretion proceeds near the BHL rate, the BH consumes the dense core within hours and the star is destroyed in an engine-powered, collapsar-like transient---with a long gamma-ray burst should the jet break out of the envelope; the visibility of such an explosive transient depends on its location in the AGN disk \citep{Perna:2021}. If instead feedback caps the accretion luminosity near the BH's Eddington value, $L_{\rm Edd,BH}\simeq1.5\times10^{39}\,(m_{\rm BH}/10\,\Msun)$~erg/s, the central engine supplies only $\sim1\%$ of the star's own near-Eddington luminosity ($L_{\ast}\sim10^{41}$~erg/s). The BH then acts as a weak central energy source, similar to a burning shell but with a much higher mass-to-energy efficiency ($\eta\sim0.1$), and the configuration is a quasi-star \citep{Begelman08,BallTout11}. In this regulated regime the star is not destroyed on any short timescale: injecting the net binding energy takes $\sim0.6$~Myr, while the immortal re-gains mass from the disk $\sim10^{4}$ times faster than the BH grows at its Eddington rate. Notably, the fiducial configuration ($m_{\rm BH}/M_{\ast}\simeq1.3\%$) sits just below the maximum core-to-envelope mass ratio ($\sim2\%$) beyond which hydrostatic quasi-star solutions cease to exist \citep{BallTout11}, so both a long-lived quasi-star phase and eventual envelope dispersal are plausible endpoints.

Figure~\ref{fig:power} compares the magnitude of the dissipative channels in Eq.~\ref{eq:dLdt} directly. A single BH sinking through the stellar potential also emits gravitational waves, since the system has a time-varying mass quadrupole, but this contribution is dynamically negligible compared with gas drag. For comparison with that channel, we instead evaluate the GW power for a representative binary, taking the sequential-capture geometry (\S\ref{sec:bbh}) as a concrete example: a $10\,\Msun$ BH (BH2) spiraling toward a second, equal-mass BH (BH1) left near the stellar center by an earlier capture. The relevant speed is then the binary's relative orbital velocity $v_{\rm k}(a)=\sqrt{G\,[m_1+m_2+M_{\rm en}(a)]/a}$, which interpolates between the stellar sinking speed $\sqrt{G M_{\rm en}/a}$ in the envelope and the BBH relative velocity $\sqrt{G(m_1+m_2)/a}$ in the dense core. Across essentially the entire envelope the two gas-drag channels exceed the GW power by many orders of magnitude, GW emission overtaking them only in the final approach to merger; for a single BH sinking through the otherwise BH-free interior the GW channel is utterly negligible, and in either case the orbit decays through gas drag alone. This same comparison frames the internal binary-hardening problem that sets in once the pair reaches the dense core (\S\ref{sec:bbh}).

\subsection{AGN Star + BBH}
\label{sec:bbh}
Before developing the binary scenario we must address a tension with \S\ref{sec:orbital}, where a single sunk BH can destroy the immortal through feedback-limited accretion. A sequential channel (a first BH sinks, a second arrives later) only operates if the star survives long enough between encounters, i.e. if the single-BH destruction time exceeds the mean time between encounters, $t_{\rm destroy,1BH}\gtrsim\Gamma_{\ast,\bullet}^{-1}\sim5\times10^{4}\,{\rm yr}$ (\S\ref{sec:rates}). The destruction time is bracketed by two limits. If accretion proceeds near the Bondi--Hoyle--Lyttleton rate, which is hyper-Eddington at the core density, the star is consumed in $\ll\Gamma_{\ast,\bullet}^{-1}$ and the single BH destroys the immortal before a companion arrives. If instead feedback regulates the accretion to near the Eddington rate, delivering the star-unbinding energy at an escaping luminosity $\lesssim L_{\rm Edd,BH}$ takes $\gtrsim$\,Myr$\,\gg\Gamma_{\ast,\bullet}^{-1}$, and the star survives long enough to capture a second BH. (Given the photon trapping discussed in \S\ref{sec:perturb}, the regulator must be mechanical, i.e.\ disk winds and jets; Eddington serves here only as a reference scale.) In this regulated regime the star is effectively a quasi-star whose central engine supplies only a percent of the stellar luminosity (\S\ref{sec:orbital}), so survival is the natural outcome rather than a fine-tuned one. We therefore distinguish two routes to an embedded BBH: \emph{(i) sequential capture}, available only when the survival inequality holds, in which a first BH sinks, the star survives feedback-limited accretion, and a second BH arrives; and \emph{(ii) direct capture}, in which a pre-existing or dynamically-assembled pair enters the star before any single central BH can destroy it. In the fast-destruction limit only the direct route survives, and the channel is then gated by the (uncertain) binary fraction among the compact objects the immortal encounters. The hardening physics below applies to both; only the sequential route is conditional on the survival inequality, and which limit obtains is itself an open question requiring radiation-hydrodynamic models of feedback-regulated accretion onto a core BH. The two channels differ in the approach: a test particle sinking onto a central BH (sequential) versus a binary whose center of mass sinks while its internal orbit is preserved (direct, with the heavier pair sinking as a single perturber). Both deliver a hard internal binary, $a_{\rm BBH}\lesssim R_{\odot}$, into the dense core. The hardening that sets the merger time and GW imprint acts on that internal orbit, where the two BHs orbit their common center of mass at $v_{\rm rel}=\sqrt{G M_{\rm BBH}/a_{\rm BBH}}$, so the result is insensitive to which channel formed the pair.

A second BH that sinks to the center of an immortal within $\lesssim1\,{\rm yr}$ immediately forms a very hard (tightly bound) BBH.
The mutual Hill sphere of a $M_{\ast} \sim 10^{2}M_{\odot}$ plus a $20M_{\odot}$ BBH can be parameterized by
\begin{equation}
    R_{{\rm H},\ast, {\rm BBH}} = 7r_{g} \left(\frac{R_{\rm AGN}}{10^{3}r_{g}}\right) \left(\frac{M_{\ast,{\rm BBH}}}{120M_{\odot}} \right)^{1/3} \left( \frac{M_{\rm SMBH}}{10^{8}M_{\odot}}\right)^{-1/3}
\end{equation}
where $R_{\rm AGN}$ is the binary's orbital radius about the SMBH, $M_{\ast,{\rm BBH}}=M_{\ast}+M_{\rm BBH}$, and $r_{g}=GM_{\rm SMBH}/c^{2} \sim 1\,{\rm au}\,M_{8}$ with $M_{8}=M_{\rm SMBH}/10^{8}M_{\odot}$. Thus, for a BBH at $R_{\rm AGN}\simeq10^{3}r_{g}$ around a $10^{8}M_{\odot}$ SMBH, $R_{{\rm H},\ast,{\rm BBH}}\simeq7r_{g}\simeq7\,{\rm au}\simeq1500\,R_{\odot}$; at fixed $R_{\rm AGN}/r_{g}$ the Hill radius scales as $R_{{\rm H},\ast,{\rm BBH}}\propto M_{\rm SMBH}^{2/3}$ (since $r_{g}\propto M_{\rm SMBH}$ while the Hill factor contributes $M_{\rm SMBH}^{-1/3}$). From our results above, two BHs will rapidly come close to a separation $a_{\rm BBH}<R_{\odot}$ and form a binary. The resulting BBH has binding energy 
\begin{eqnarray}
  E_{\rm BBH} &=& -\frac{Gm_{1}m_{2}}{2a_{\rm BBH}}\\
  &\simeq& -1.9 \times 10^{50}{\rm erg}\left(\frac{m_{1}}{10M_{\odot}}\right)\left(\frac{m_{2}}{10M_{\odot}}\right)\left(\frac{a_{\rm BBH}}{R_{\odot}}\right)^{-1}\nonumber
\end{eqnarray}
which is hard with respect to any perturber population of velocity dispersion $\sigma < v_{\rm orb} \simeq 2000~{\rm km/s}$\footnote{The orbital velocity of the binary system is $v_{\rm orb} \simeq 2000~{\rm km/s}
\left(M_{\rm BBH}/20\,M_{\odot}\right)^{1/2}
\left(a_{\rm BBH}/R_{\odot}\right)^{-1/2}$.}, a condition comfortably met both by disk-embedded BHs encountered within the Hill sphere ($\mathcal{O}(50\,{\rm km/s})$) and by the wider nuclear cluster ($\sigma\sim10^{2}\,{\rm km/s}$). 
Note that at fixed mass ratio $E_{\rm BBH}\propto M_{\rm BBH}^{2}$, so the binding energy grows steeply for more massive binaries---a point we return to in \S\ref{sec:noimmortal}.

\begin{figure}
\centering
\includegraphics[width=\columnwidth]{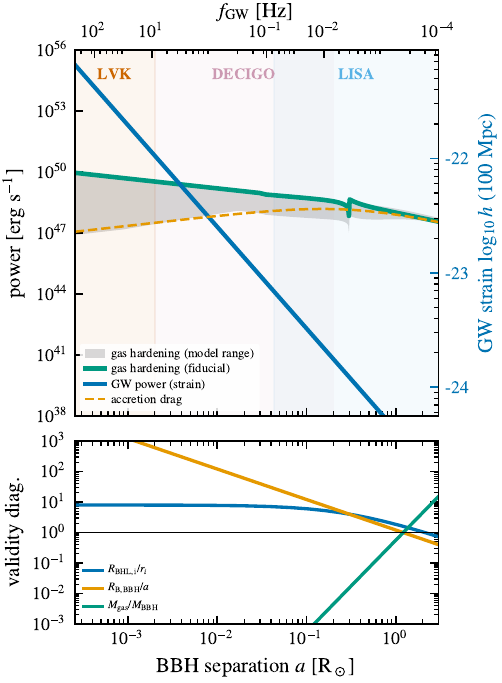}
\caption{Energy-loss channels for the internal BBH orbit, the analog of Fig.~\ref{fig:power}, as a function of internal separation $a_{\rm BBH}$. The gray band spans the range of gas-hardening prescriptions and the solid line is the fiducial model (nonlinear self-wake plus the \citet{KKSS08} companion wake); the GW power (with the corresponding characteristic strain at $d=100$~Mpc on the right axis) and the accretion drag are also shown. The top axis gives the GW frequency, with the LISA, DECIGO, and LVK bands shaded. The lower panel shows validity diagnostics---the per-BH Bondi radius relative to its orbital radius $R_{\rm BHL,i}/r_{i}$, the binary Bondi radius relative to the separation $R_{\rm B,BBH}/a$, and the enclosed gas mass relative to the BBH mass; the independent point-perturber treatment breaks down where these exceed unity.}
\label{fig:powerB}
\end{figure}

Whether hardening truly stalls in this regime is the central uncertainty of the problem. Once the binary's center of mass has reached the dense core, its internal orbit keeps losing energy to the same gas channels---dynamical friction and accretion drag---now competing with GW emission. Figure~\ref{fig:powerB} compares these channels along the internal inspiral: gas drag dominates the hardening from $a_{\rm BBH}\sim R_{\odot}$ through the deci-Hz regime, exceeding the GW luminosity by many orders of magnitude over the separations that set the residence time, while GW emission takes over only close to merger. The corresponding semi-analytic trajectory for this sequential-capture configuration (whose power budget is shown in Fig.~\ref{fig:power}) is plotted in Fig.~\ref{fig:spiral}(b): the pair winds $\sim\!3000$ times, crowding into the gas--GW crossover before plunging to the ISCO. Linear theory \citep{Ostriker99,Kim207} predicts the drag to weaken once the orbital motion turns supersonic, whereas the nonlinear circular-orbit treatment of \citet{Kim10}, supplemented by the double-perturber wake of \citet{KKSS08}, keeps the binary hardening through the supersonic regime.

We bracket this uncertainty with a ladder of hardening prescriptions (Appendix~\ref{apndx:ladder}, Fig.~\ref{fig:ladder}), spanning the linear (near-stalling) limit, the nonlinear self-wake, the companion double-wake, and common-envelope wind-tunnel drag coefficients. Integrating the hardening rate, our fiducial model (nonlinear self-wake plus companion wake) drives the binary from $a_{\rm BBH}\sim R_{\odot}$ to merger in $\sim1\,{\rm hr}$; the physically motivated rungs span $\sim1$--$8\,{\rm hr}$, and every prescription drives merger at least $10^{4}$ times faster than the corresponding vacuum (GW-only) inspiral. Only the pure-linear rung, which we treat as a cautionary lower-drag limit, approaches a stall (yr-scale). Unless stated otherwise the ladder uses the gas-only (ideal-gas) sound speed; adopting instead the radiation-dominated adiabatic sound speed ($\beta\equiv P_{\rm gas}/P_{\rm tot}\simeq0.2$ in the core, so $c_{s}$ is $\sim\!2\times$ larger) lengthens the fiducial merger to $\sim\!4\,{\rm hr}$ and the band to $\sim\!3$--$20\,{\rm hr}$, but does not change the qualitative conclusion---merger in hours rather than the $\sim\!10^{5}\,{\rm yr}$ vacuum time.

\begin{figure}
\centering
\includegraphics[width=\columnwidth]{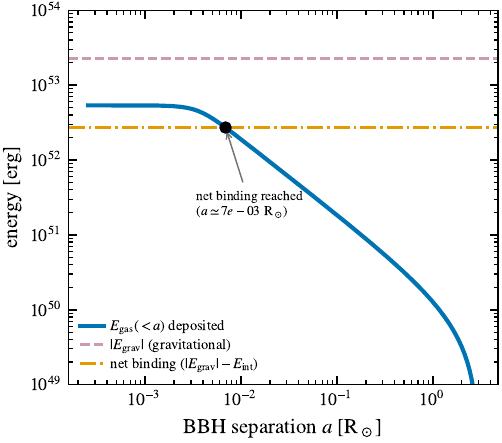}
\caption{Cumulative orbital energy delivered to the stellar gas as the fiducial $10+10\Msun$ binary hardens to internal separation $a$, $E_{\rm gas}(<a)=\int L_{\rm gas}\,dt$, compared with the gravitational binding energy $|E_{\rm grav}|$ of the immortal and its net binding energy $|E_{\rm grav}|-E_{\rm int}$ (reduced by the internal energy of the radiation-dominated, near-$n{=}3$ envelope). The deposited energy reaches the net binding energy near $a\simeq7\times10^{-3}R_{\odot}$, indicating that even a fiducial-mass BBH can deposit a dynamically significant amount of energy into the core before GW emission takes over.}
\label{fig:edep}
\end{figure}

These timescales are necessarily approximate. Across the supersonic inspiral the two BHs are not independent point perturbers: their individual Bondi radii exceed their orbital radii ($R_{\rm BHL,i}/r_{i}\gtrsim\,$few; lower panel of Fig.~\ref{fig:powerB} and the pink-shaded core of Fig.~\ref{fig:spiral}b). The deep core is therefore a coupled, nonlinear binary--gas problem that only a 3D calculation including self-gravity can resolve. The ladder also shares a systematic that no rung captures. Every prescription assumes an unperturbed, continuously replenished gas background, whereas the energy (Fig.~\ref{fig:edep}) and angular momentum deposited by the binary exceed what the local core gas holds. The heated, spun-up gas is expected to provide less drag, so the band should be read as an upper envelope on the drag rather than a two-sided error bar. The merger itself is nevertheless robust: by the time the local gas could be strongly disturbed, the binary is already compact enough that vacuum GW emission alone completes the inspiral within hours (Eq.~\ref{eq:tgw} below). The resulting $\mathcal{O}(1)$ imprint on the emitted waveform is discussed in \S\ref{sec:gwobs}.

All the gas-hardened predictions that follow use this same fiducial prescription: the merger times quoted here, the core energy deposition (Fig.~\ref{fig:edep}), and the dephasing and characteristic strain (Figs.~\ref{fig:dephasing} and~\ref{fig:hstrain}). The fiducial model combines the per-component nonlinear self-wake of \citet{Kim10} with the \citet{KKSS08} companion wake (Model~2 of Appendix~\ref{apndx:ladder}), evaluated at the core density and sound speed. It uses the same nonlinear dynamical-friction kernel as the single-perturber estimate of Fig.~\ref{fig:power}, but is evaluated component by component with the companion term retained (a lone sinking BH has no companion wake).

The rapidity of this hardening raises a back-reaction concern for the fiducial $10+10\Msun$ binary itself, not only for the massive binaries of \S\ref{sec:noimmortal}. Because gas drag absorbs almost all of the orbital energy released until GW emission takes over, the energy delivered to the surrounding stellar gas as the binary hardens from its capture separation $a_{0}$ to separation $a$ is $E_{\rm gas}(<a)\simeq E_{\rm BBH}(a)-E_{\rm BBH}(a_{0})\simeq Gm_{1}m_{2}/2a$ for $a\ll a_{0}$. Figure~\ref{fig:edep} compares this cumulative deposition with the star's binding energies. By $a_{\rm BBH}\sim10^{-2}R_{\odot}$ the deposited energy reaches $\sim2\times10^{52}\,$erg---of order $10\%$ of the gravitational binding energy $|E_{\rm grav}|\simeq2.3\times10^{53}\,$erg, and comparable to the net binding energy ($|E_{\rm grav}|-E_{\rm int}\simeq2.7\times10^{52}\,$erg) of the radiation-dominated envelope, which it reaches near $a\simeq7\times10^{-3}R_{\odot}$. The deposition is therefore not a negligible perturbation even at fiducial mass.

Moreover, it occurs over the $\sim1\,$hr hardening time, comparable to the stellar dynamical time $\sqrt{R_{\ast}^{3}/GM_{\ast}}\approx0.7\,$hr, so the background star cannot be assumed static during the hardening episode. A fully consistent treatment of the deep core therefore requires the gas response and the stellar back-reaction to be evolved together, reinforcing that these timescales are merely suggestive. However, this does not prevent the merger itself: by the time enough energy has been deposited to disturb the core ($a\sim10^{-2}R_{\odot}$), the binary would already be tight enough that vacuum GW emission alone (Eq.~\ref{eq:tgw}) would complete the merger in $\sim7\,$hr---and in $\sim1.7\,$hr by $a\simeq7\times10^{-3}R_{\odot}$. Stellar back-reaction can therefore alter the environmental phase imprint, but is unlikely to prevent the merger once the binary has reached the dense core.

In the absence of gas, the time for a circularized equal-mass BBH of mass $M_{\rm BBH}=m_{1}+m_{2}=2m_{1}$ and separation $a_{\rm BBH}$ to merge is given by \citep{Peters64}
\begin{equation}
    t_{\rm GW} = \frac{5}{256} \frac{c^{5}}{G^{3}} \frac{a_{\rm BBH}^{4}(1-e_{\rm BBH}^{2})^{7/2}}{m_{1}m_{2}M_{\rm BBH}}
\end{equation}
which, assuming the BBH has circularized around its own center of mass ($e_{\rm BBH} \sim 0$), can be parameterized as
\begin{equation}
    t_{\rm GW} \sim 0.08\,{\rm Myr} \left( \frac{a_{\rm BBH}}{1R_{\odot}}\right)^{4} \left( \frac{m_{1}}{10M_{\odot}}\right)^{-3}\,.
    \label{eq:tgw}
\end{equation}
An immortal star can encounter BHs at a rate up to $\Gamma_{\ast,\bullet}\sim20\,{\rm Myr}^{-1}$ ($\Gamma_{\ast,\bullet}^{-1}\sim5\times10^{4}\,{\rm yr}$; \S\ref{sec:rates}), so this vacuum merger time at $a_{\rm BBH}\sim R_{\odot}$ is comparable to the (optimistic) time to encounter an additional embedded BH. If gas hardening were inefficient, such a tertiary encounter (arriving at the Keplerian differential velocity of $\mathcal{O}(50\,{\rm km/s})$ within the mutual Hill sphere) could in principle harden the binary further \citep{Leigh18}. Note, however, that a tertiary penetrating the stellar envelope is itself captured by gas drag within $\sim0.1$\,yr (\S\ref{sec:orbital}), so the interaction would be a dissipative triple rather than a clean three-body flyby. In our fiducial gas-hardened core the binary merges in hours, long before another encounter is expected, so the merger is gas-driven rather than encounter-driven.

At merger, the remnant BH receives a recoil kick $v_{\rm kick}$, whose magnitude depends on the mass ratio $q=m_{2}/m_{1}$, the spins $\chi_{1},\chi_{2}$, and the spin--orbit alignments of the progenitors \citep[e.g.][]{Varma22}. The escape velocity from an immortal is $v_{\rm esc}=\sqrt{2GM_{\ast}/R_{\ast}} \sim 1950\,{\rm km/s}\,(M_{\ast}/10^{2}M_{\odot})^{1/2}(R_{\ast}/10R_{\odot})^{-1/2}$. A remnant kicked with $v_{\rm kick} < v_{\rm esc}$ is displaced but damped by the same gas drag that drove the inspiral, and quickly re-settles toward the core (to the extent that the post-merger star retains its dense interior; \S\ref{sec:noimmortal}). Kicks are maximal for moderately asymmetric mass ratios ($q\sim0.3$--$0.4$) and for comparable-mass, rapidly spinning, spin-misaligned binaries, and vanish in the extreme-mass-ratio limit \citep{Varma22,IslamWadekar26}. The prograde, spin-aligned bias expected in this channel (\S\ref{sec:bbh_props}) and the decreasing mass ratio of any growing primary both suppress the largest kicks, favoring retention. By contrast, isotropically oriented spins, as expected in gas-free dynamical channels, make superkick configurations accessible and escape considerably more likely \citep{Borchers25}.

Merger also radiates $\sim5\%$ of the binary rest mass in GWs, $E_{\rm GW}\sim2\times10^{54}$\,erg for the fiducial binary. This energy leaves the star without depositing heat, since stellar material is essentially transparent to GWs. The accompanying sudden shallowing of the central potential does perturb the gas bound to the binary, but the associated energy change, $\sim0.05\,G M_{\rm BBH} M_{\rm gas}/r \sim 10^{50}$\,erg for the $\sim20\,\Msun$ of gas within $\sim R_{\odot}$, is far below the hardening deposit of Fig.~\ref{fig:edep} and produces no qualitative change.

As the mass of the merged BH grows, the binding energy of subsequent binaries grows as $M_{\rm BBH}^{2}$, and with it the energy deposited in the star during hardening. Sufficiently massive binaries perturb and ultimately unbind the star, leaving behind a hard, massive BBH (\S\ref{sec:noimmortal}).

\subsection{BBH properties}
\label{sec:bbh_props}
The properties of a BBH inside an immortal depend on whether the secondary arrives prograde or retrograde with respect to the immortal spin. An immortal $\mathcal{O}(100M_{\odot})$ has a Type I migration timescale an order of magnitude faster than an isolated BH $\mathcal{O}(10\,M_{\odot})$ (see Eq.~\ref{eq:mig} below). Thus, in the frame of the immortal, encounters should typically occur as the immortal `catches up' with a BH on an interior orbit. For a sufficiently close impact parameter ($\Delta R$), such that the BH is effectively co-orbital, the encounter arrives at the local Keplerian shear velocity of $\mathcal{O}(50\,{\rm km/s})$. The capture geometry then favors, though does not guarantee, a prograde entry: simulations of gas-assisted binary formation in AGN disks find both senses of rotation, with the outcome depending on the impact parameter and the gas response \citep{DeLaurentiis23,Rowan23,Qian24,Dodici24}.

A retrograde entry changes the details but probably not the outcome. The relative velocity between a retrograde orbiter and the envelope gas is larger, and the supersonic dynamical friction correspondingly weaker \citep[e.g.][]{Ostriker99}; but the linear suppression is a factor of a few to ten, so the sink time lengthens from $\sim0.1\,$yr to at most years---still essentially instantaneous compared with every other timescale in the problem (the encounter interval, accretion-driven realignment, and migration). Dynamical friction on a retrograde orbiter also pumps its orbital eccentricity rather than damping it, as found for single retrograde orbiters in AGN disks \citep{Secunda21} and for retrograde binaries \citep{Calcino24,DDL25}. For the sinking object this raises the pericenter drag and shortens the sink further. For a retrograde internal binary it opens an additional route to coalescence: runaway eccentricity growth can bring the pair close enough at pericenter for GW emission to take over \citep{Calcino24}, and such systems may reach the LVK band with detectable eccentricity (\S\ref{sec:gwobs}). We therefore do not expect a long-lived population of wide, stalled pairs. Nor would such a pair be a two-body binary in any dynamical sense: at $a_{\rm BBH}\sim10\,R_{\odot}$ inside a star of radius $12.5\,R_{\odot}$ the enclosed stellar mass ($M_{\rm en}\sim500\,M_{\odot}\gg M_{\rm BBH}$) dominates the potential, the two BHs independently orbit the stellar center at $\sim\!4000\,{\rm km/s}$, and external perturbers cannot ionize a configuration whose binding is dominated by the stellar gas. Accretion torques can also realign a retrograde pair toward prograde over time \citep{Alex24}, although at the feedback-limited accretion rates required for the star's survival (\S\ref{sec:bbh}) this realignment is slow, so the prograde bias rests mainly on the entry geometry.

The net expectation is thus a bias, not a guarantee, toward prograde, rapidly hardening binaries, with retrograde entries mainly delaying the hardening and reducing the net accreted angular momentum. This has consequences for gravitational-wave observables (\S\ref{sec:discussion}). The embedded BHs also accrete and are torqued by the surrounding gas; for prograde, disky accretion this tends to align their spins with the immortal's rotation axis, although the mass any component can accrete during its short residence time is limited by the same feedback bracket of \S\ref{sec:bbh}. The fiducial gas-hardened binary merges in hours, long before a subsequent dynamical encounter (\S\ref{sec:bbh}).

\subsection{No longer immortal}
\label{sec:noimmortal}
Finally, we ask how massive an embedded binary can become before it destroys its host. The energy delivered to the stellar gas during hardening is the binary binding energy evaluated where GW emission takes over, i.e.\ at the gas--GW crossover ($a_{\rm x}\sim{\rm few}\times10^{-3}\,R_{\odot}$ for the fiducial binary). At fixed mass ratio this energy grows as $M_{\rm BBH}^{2}$. The relevant threshold for ejecting stellar material is the \emph{net} envelope binding energy $E_{\ast}=|E_{\rm grav}|-E_{\rm int}$ (using $|E_{\rm grav}|$ instead gives a conservative upper limit). By this criterion even the fiducial $10+10\,\Msun$ binary is disruptive: Fig.~\ref{fig:edep} shows its cumulative deposition reaching $E_{\ast}$ at $a\simeq7\times10^{-3}\,R_{\odot}$ and $\sim\!2E_{\ast}$ by the crossover. The maximum binary mass that can merge while leaving the star essentially intact is therefore at or below the fiducial $\sim20\,\Msun$. Evaluating $E_{\rm BBH}$ at $a_{\rm BBH}=R_{\odot}$ would instead suggest $\mathcal{O}(100$--$200\,M_{\odot})$, but the hardening binary passes through that separation in minutes, having deposited little energy by then. Whether the deposited energy actually unbinds the envelope depends on how efficiently it couples to the bulk of the star within the hours-long episode, a question only 3D simulations can settle; ``destruction'' should be read as ``heated by of order its net binding energy.'' The merger itself is not prevented: by the time $E_{\ast}$ has been deposited, vacuum GW emission alone completes the coalescence within hours (Eq.~\ref{eq:tgw}). Each processed immortal is thus left strongly heated or unbound, becoming a localized source of mixing and turbulence in the AGN disk. The channel therefore yields only one to a few mergers per star, consistent with the yield adopted in \S\ref{sec:rates}. Extended hierarchical growth to IMBH masses ($>100\,M_{\odot}$) inside a single immortal is correspondingly disfavored: each merger requires a fresh (or re-formed) host, and LVK observations may thereby indirectly constrain immortal-star masses, lifetimes, and resupply.

\section{Rates}
\label{sec:rates}
To a first approximation, the encounter rate between an AGN star at initial radius $r_{0}$ and BHs on interior orbits is
\begin{equation}
    \Gamma_{\ast,\bullet} \approx \frac{N_{\bullet}(<r_{0})}{\tau_{\rm mig}(r_{0})}
\end{equation}
where $N_{\bullet}(<r_{0})$ is the number of embedded BHs at disk radii $<r_{0}$ and $\tau_{\rm mig}(r_{0})$ is the migration timescale of an immortal located at $r_{0}$ in the disk. We assume that the immortal star mass $M_{\ast} \gg M_{\bullet}$ such that we can ignore BH migration during $\tau_{\rm mig}$. In this scenario $\Gamma_{\ast, \bullet}$ should be highest for the first immortals that migrate quickly inwards and gather the largest number of BHs on interior orbits.

The Type I migration timescale for an embedded mass (immortal star or BH) is adapted from \citet{2004ApJ...602..388T} as
\begin{equation}
    \tau_{\rm mig} = \frac{1}{N} \frac{M_{\rm SMBH}}{q\Sigma_{0}r_{0}^{2}} \frac{h^{2}}{\Omega}
\end{equation}
where $N \sim 3$ is a numerical factor, $M_{\rm SMBH}$ is the SMBH mass, $q=m/M_{\rm SMBH}$ is the mass ratio of the orbiter mass ($m$) to $M_{\rm SMBH}$, $\Sigma_{0}$ is the disk surface density at the initial position ($r_{0}$) of the orbiter, $h$ is the disk aspect ratio and $\Omega$ the orbital frequency of the orbiter. The migration timescale can be parameterized as
\begin{eqnarray}
    \tau_{\rm mig, \ast} &\approx& 0.5\,{\rm Myr} \left(\frac{N}{3}\right)^{-1} \left(\frac{r_{0}}{10^{4}r_{g}}\right)^{-1/2} \left(\frac{M_{\ast}}{10^{2}M_{\odot}}\right)^{-1} \nonumber \\
    &\times& \left(\frac{h}{10^{-3}}\right)^{2}\left(\frac{\Sigma_{0}}{10^{2}{\rm g/cm^{2}}}\right)^{-1}\left(\frac{M_{\rm SMBH}}{10^{8}M_{\odot}}\right)
    \label{eq:mig}
\end{eqnarray}
where $r_{g}=GM_{\rm SMBH}/c^{2}$ and the parameter values are appropriate to the disk model of \citet{2005ApJ...630..167T}. The equivalent timescale is shorter for the higher-surface-density disk model of \citet{Sirko03}. Thus, immortal stars should migrate onto the SMBH well within fiducial AGN lifetimes of several Myr. Complications that may change this scenario include: the effect of radiative feedback \citep{Hankla20} or turbulence \citep{Laughlin04,McPike26,Garg26} on migration gas torques, or stalling at migration traps/swamps \citep{Bellovary16,Grishin24,McFACTS25}, or indeed, a shorter than expected disk lifetime. Nevertheless, if we assume an in-migration of immortals given by $\tau_{\rm mig}$, then we can estimate the average encounter rate between an immortal star and embedded BH.

Nuclear star clusters (NSCs) do not continue to increase in mass for $M_{\rm SMBH} \geq 10^{7}$--$10^{8}M_{\odot}$ \citep{Neumayer20}. If we assume a Milky Way--like NSC, of mass $M_{\rm NSC} \sim 3 \times 10^{7}M_{\odot}$, with a Bahcall--Wolf number density profile $n_{\rm NSC} \propto r^{-7/4}$ in the central $\sim 0.25\,{\rm pc}$, then there are $N_{\bullet} \sim \mathcal{O}(10^{4})$ BHs within the central parsec \citep[e.g.][]{Generozov18}. For an AGN disk of size $R_{\rm disk} \sim 0.2\,{\rm pc}$ (i.e. $\simeq5\times10^{4}\,r_{g}$ for $M_{8}=1$, where $M_{8}=M_{\rm SMBH}/10^{8}M_{\odot}$), with average aspect ratio $\bar{h} \sim 0.03$, we find
an initial embedded population of $N_{\bullet} \sim 200\,(R_{\rm disk}/0.2\,{\rm pc})$. For a uniform distribution of BHs throughout the disk, $N_{\bullet}(r_{0}<10^{4}r_{g}) \sim 10$, so $\Gamma_{\ast,\bullet} \sim 20\,{\rm Myr}^{-1}$ for the fiducial immortal of Eq.~(\ref{eq:mig}). This is a conservative estimate, since it does not account for the AGN stars' contribution to the BH population \citep[e.g.][]{Cantiello:2021}.

If we assume that all BHs on interior disk orbits are captured by this immortal and driven through a series of mergers during the time $\tau_{\rm mig}$, the corresponding merger yield would be, in the optimistic limit, $n_{\rm BBH} \sim 20$ per AGN per Myr. Given the energy-deposition results of \S\ref{sec:noimmortal}, however, each immortal likely survives only one or two mergers, so the realized yield is closer to $n_{\rm BBH}\sim1$--$2$ per AGN per Myr. Sustaining even this reduced yield over an AGN lifetime requires a continuing supply of immortals (by in-situ formation, capture, and regrowth; \citealt{Cantiello:2021,Fabj:2025,Dittmann:2025}) and of embedded BHs, which we do not model here. The number density of galaxies with mass greater than or equal to the Milky Way is $n_{\rm GN} \sim 4 \times 10^{-3}{\rm Mpc^{-3}}$ \citep{Baldry12}. Assuming an AGN fraction $f_{\rm AGN} \sim 0.1$ yields a BBH merger rate of
\begin{eqnarray}
    \mathcal{R} &\sim& 8 {\rm Gpc}^{-3} {\rm yr}^{-1} \nonumber \\
    &\times& \left(\frac{n_{\rm GN}}{0.004 {\rm Mpc}^{-3}}\right)\left(\frac{f_{\rm AGN}}{0.1}\right) \left(\frac{n_{\rm BBH}}{20/{\rm AGN/Myr}}\right)
\end{eqnarray}
which, for the optimistic normalization $n_{\rm BBH}=20$, is $\sim 1/3$ of the O3-era LVK merger rate $\mathcal{R} \sim 24\,{\rm Gpc}^{-3}{\rm yr}^{-1}$ \citep{Abbott21}; the more likely $n_{\rm BBH}\sim1$--$2$ instead gives $\mathcal{R}\sim0.4$--$0.8\,{\rm Gpc}^{-3}{\rm yr}^{-1}$. We quote $\mathcal{R}\lesssim8\,{\rm Gpc}^{-3}{\rm yr}^{-1}$ as an optimistic ceiling rather than a two-sided range, because the dominant uncertainty acts only downward. The fiducial migration time (Eq.~\ref{eq:mig}) is normalized to a thin disk region with $h=10^{-3}$, while the population estimate uses the disk-averaged aspect ratio $\bar h\sim0.03$; since $\tau_{\rm mig}\propto h^{2}$, the two differ by a factor $\sim\!10^{3}$. If immortals do not spend their migration in thin, fast-migrating parts of the disk, the rate falls by a corresponding factor, plausibly to $\ll1\,{\rm Gpc}^{-3}{\rm yr}^{-1}$. Thus, if immortal stars form efficiently, are steadily resupplied, and migrate through thin disk regions, they could contribute substantially to the AGN-channel BBH merger rate; the estimate is best read as an order-of-magnitude ceiling rather than a prediction.

\section{Discussion}\label{sec:discussion}

\subsection{Observables: GW}
\label{sec:gwobs}
\begin{figure*}
\centering
\includegraphics[width=\linewidth]{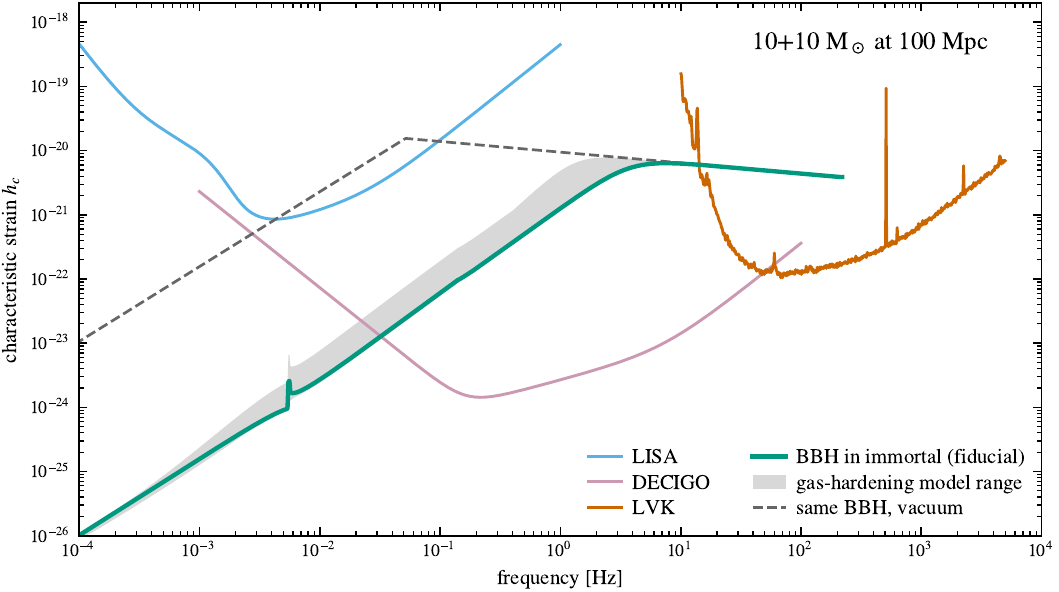}
\caption{Characteristic strain $h_{\rm c}(f)$ (Eq.~\ref{eq:hc}, computed with $T_{\rm obs}=4$~yr) of a gas-hardened $10+10\Msun$ binary at $100$~Mpc. The solid curve is the fiducial model and the shaded band spans the ladder of hardening prescriptions; the dashed curve is the same binary inspiraling in vacuum, shown against the sky-averaged LISA, DECIGO, and LVK sensitivity curves \citep{Robson19,YagiSeto11,Abbott21}. The finite observation time limits the slowly-evolving vacuum source in the milli-Hz band rather than letting it accumulate its formal full-inspiral strain. Gas hardening erases the milli-Hz inspiral cycles, suppressing $h_{\rm c}$ by several orders of magnitude in the LISA band; the track recovers toward the vacuum value at deci-Hz and continues into the LVK band as the binary approaches merger. The source is therefore a deci-Hz/DECIGO-band phenomenon whose late inspiral enters the LVK band like a standard merger.}
\label{fig:hstrain}
\end{figure*}

We have shown that immortal stars can rapidly form hard BBHs of separation $\sim R_{\odot}$. The GW frequency of such binaries can be parameterized as
\begin{equation}
 f_{\rm GW} \sim 1\,{\rm mHz} \left( \frac{M_{\rm BBH}}{20M_{\odot}}\right)^{1/2} \left(\frac{a_{\rm BBH}}{R_{\odot}}\right)^{-3/2}
\end{equation}
which lies in the LISA band \citep{LISA23}. Even in the limit most favorable to LISA, in which we ignore the gas-driven evolution and treat the binary as monochromatic over an observation time of $1$~yr ($\dot{f}_{\rm GW} \simeq 0$), the characteristic strain is only
\begin{equation}
h_{\rm char}^{2} \sim \left(\frac{N}{8}\right) h_{0}^{2}
\end{equation}
where $h_{0}$ is the binary strain
\begin{equation}
  h_{0}=r_{\rm g,BBH}\left(\frac{v_{\rm BBH}}{c}\right)^{2}\left(\frac{1}{D}\right)
\end{equation}
and $r_{\rm g,BBH}=GM_{\rm BBH}/c^{2}$ is the BBH gravitational radius, $v_{\rm BBH}$ is the binary's relative orbital velocity and $D$ is the distance to the BBH. $h_{\rm char}$ is averaged over all angles and $N \sim f_{\rm GW} \times 1\,{\rm yr}$. Even for a relatively nearby AGN ($D \sim 100\,$Mpc),
\begin{eqnarray}
    h_{\rm char} &\sim& 3\times10^{-23} \left( \frac{f_{\rm GW}}{1\,{\rm mHz}}\right)^{1/2}\left(\frac{D}{100\,{\rm Mpc}}\right)^{-1} \left(\frac{M_{\rm BBH}}{20M_{\odot}} \right) \nonumber \\
    &\times& \left( \frac{v_{\rm BBH}}{2000\,{\rm km/s}}\right)^{2}
\end{eqnarray}
which is not directly detectable by LISA, since we require $h_{\rm char}>10^{-20}$ to be resolved over the Galactic background at $\sim 1\,$mHz \citep{LISA23}. The actual gas-hardened binary sweeps through the milli-Hz band in hours rather than persisting for $N$ cycles (\S\ref{sec:bbh}), so this monochromatic estimate is a strict upper limit; the full calculation below (Eq.~\ref{eq:hc}, Fig.~\ref{fig:hstrain}) gives a far smaller LISA-band strain. By contrast, common-envelope inspirals within the Galaxy, being orders of magnitude closer, have been considered as potential LISA sources \citep{Renzo:2021}.

We expect most BBHs produced via the immortal channel in AGN to be biased to prograde (see \S\ref{sec:bbh_props}). In the fiducial picture the merger is driven by gas drag within the core, so the eccentricity evolution is set by the same uncertain nonlinear gas response: gaseous drag can damp or excite eccentricity depending on Mach number, orbital phase, and orbital sense, with retrograde configurations in particular experiencing eccentricity pumping \citep{Secunda21,Calcino24,DDL25}. If the drag circularizes efficiently, these mergers should reach the LVK band with little residual eccentricity; by contrast, retrograde systems, wide systems, or those completed by dynamical encounters may retain residual eccentricity in the LVK band \citep{Samsing22,Calcino24,DDL25}. In either case, the prograde bias implies that measurements of $\chi_{\rm eff}$ should be biased to positive values in this channel.

The positive-$\chi_{\rm eff}$ bias in this channel derives primarily from the prograde orbital bias of \S\ref{sec:bbh_props}, combined with accretion torques that tend to align BH spins with the immortal's rotation axis. We caution that the aligning torque is limited by the feedback-regulated accretion budget (\S\ref{sec:bbh}). At near-Eddington rates, the mass accreted during the first BH's residence is a small fraction of its own mass, short of what accretion-driven spin-up or full alignment requires. Strong spin alignment would therefore have to be inherited from the BHs' formation and prior accretion history in the disk, rather than acquired inside the immortal.

A caveat to the above discussion is that it treats the immortal as a fixed background, whereas \S\ref{sec:bbh} and Fig.~\ref{fig:edep} show the back-reaction is not negligible even at fiducial mass. If the immortal core sloshes in response to the BBH, this could perturb the orientation of the BBH with respect to the spin axis of the immortal. Whether such sloshing could harden the BBH still further is beyond the scope of the present paper.

If a BBH merges within the core of the star, a possibly interesting diagnostic could come from the effect of the high-density of the environment. While environmental effects on the waveform are generally studied with the goal of inferring true system parameters by eliminating possible sources of confusion \citep{Barausse2014}, there are also situations in which detecting environmental effects from changes to the waveform during a BBH merger can lead to potentially interesting discoveries, such as dark matter constraints \citep{Cole2023, Bertone2024}.  

As discussed in \S\ref{sec:bbh}, the rate of evolution of the BBH orbital separation $a(t)$
is modified by energy losses due to the BBH environment.  As a result, the time evolution of the GW frequency and phase of the signal
\begin{equation}
f(t) = \frac{1}{\pi}\sqrt{\frac{G M_{\rm BBH}}{a(t)^{3}}},\;\;\;\;
N(>f) = \int_{f}^{f_{\rm ISCO}} f'\,\frac{dt}{df'}\,{\rm d}f'
    \label{eq:nut}
\end{equation}
will differ from the vacuum case (see e.g. \citealt{Bertone2024}). Here $f(t)$ is the GW frequency, $N(>f)$ is the number of cycles accumulated from $f$ up to the innermost-stable-orbit frequency $f_{\rm ISCO}$, and $a(t)$ evolves under the combined gas and GW losses of \S\ref{sec:bbh}. The waveform experiences a dephasing with respect to the vacuum inspiral, quantified by the difference in the number of cycles from a reference frequency until merger,
\begin{equation}
\delta N(>f) = N_{\rm vac}(>f) - N_{\rm star}(>f),\;\;\;\; \delta\Phi = 2\pi\,\delta N,
\label{eq:Ncyc}
\end{equation}
where $N_{\rm vac}$ and $N_{\rm star}$ are the cycle counts for inspiral in vacuum and within the star, and $\delta\Phi$ is the corresponding accumulated phase difference in radians.

\begin{figure*}
\centering
\includegraphics[width=\linewidth]{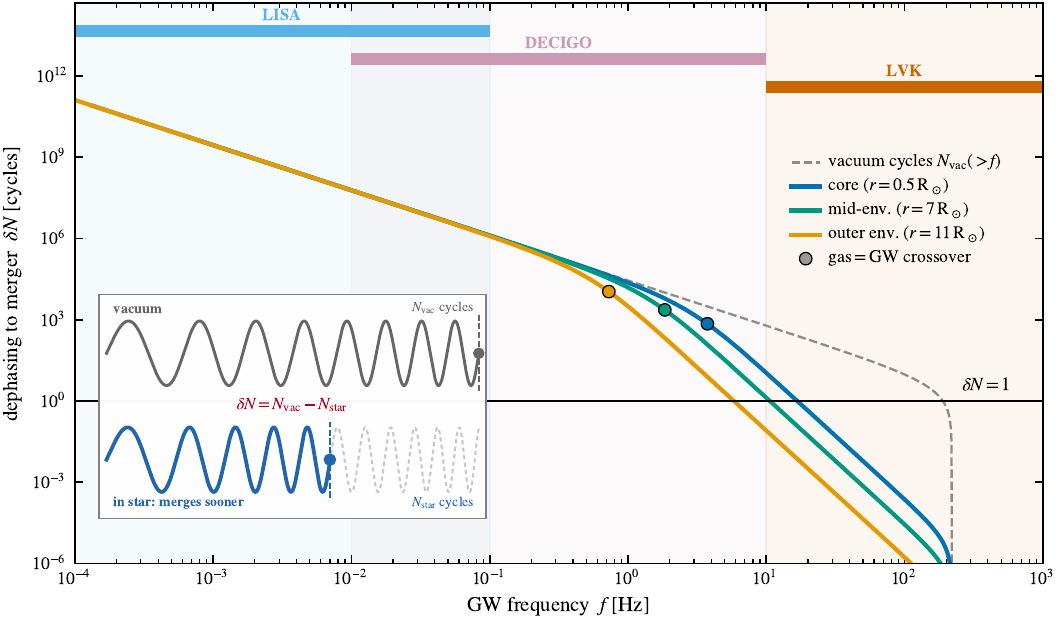}
\caption{Dephasing of the gas-hardened BBH waveform relative to a vacuum inspiral, $\delta N(>\!f)$, as a function of GW frequency, for a $10+10\Msun$ binary; the dashed gray curve is the vacuum cycle count $N_{\rm vac}(>f)$ and the colored curves are $\delta N$ for representative core gas conditions. Each colored curve is solid where the orbit is structure-robust ($a<a_{\rm unc}$, the BBH dominating the locally enclosed mass) and dotted at larger separation / lower frequency, where $M_{\rm gas}(<a)\gtrsim M_{\rm BBH}$ and the static-background approximation is uncertain. Filled circles mark the gas$=$GW crossover: to its left the binary is gas-dominated and plunges, to its right GW emission drives the inspiral. The LISA, DECIGO, and LVK bands are shaded. The milli-Hz dephasing reflects a rapid gas-driven sweep (a ``plunge'') rather than a measurable phase offset; the band-by-band interpretation and the LVK-band residual are discussed in \S\ref{sec:gwobs}. \emph{Inset:} schematic interpretation of $\delta N(f)$. Starting from the same observed GW frequency $f$, a vacuum binary completes $N_{\rm vac}$ cycles before merger, while a gas-hardened binary chirps more rapidly and merges sooner, completing only $N_{\rm star}$; their difference is the plotted dephasing $\delta N=N_{\rm vac}-N_{\rm star}$.}
\label{fig:dephasing}
\end{figure*}

It is also instructive to place the source on the strain--frequency plane through its characteristic strain $h_{\rm c}$, which folds the per-cycle GW amplitude together with the number of wave cycles spent near each frequency, so that the signal can be read directly against a detector's sensitivity curve \citep[e.g.][]{Robson19}:
\begin{equation}
h_{\rm c}(f) = A(f)\,\sqrt{N_{\rm cyc}(f)},\qquad
N_{\rm cyc}(f) = \min\!\left(\frac{f^{2}}{|\dot f|},\;\; f\,T_{\rm obs}\right),
\label{eq:hc}
\end{equation}
where $A(f) = (4/d)\,(G\mathcal{M}_{\rm c}/c^{2})\,(\pi f\,G\mathcal{M}_{\rm c}/c^{3})^{2/3}$ is the face-on (optimally oriented) strain amplitude of the circular binary ($\mathcal{M}_{\rm c}$ is the chirp mass and $d$ the luminosity distance; the rms average over sky location, inclination, and polarization is $2/5$ of this value), and $\dot f$ is the chirp rate set by the combined gas and GW losses of \S\ref{sec:bbh}. The effective number of cycles $N_{\rm cyc}$ is the lesser of the cycles the source intrinsically spends near $f$, $f^{2}/|\dot f|$, and those a detector accumulates within a finite observation time, $f\,T_{\rm obs}$ (we adopt $T_{\rm obs}=4$~yr): a rapidly-chirping binary contributes all of its near-$f$ cycles, whereas a slowly-evolving (vacuum, milli-Hz) source is capped by the mission duration rather than reaching its formal full-inspiral strain. Figure~\ref{fig:hstrain} shows $h_{\rm c}(f)$ for the fiducial gas-hardened binary.

Figure~\ref{fig:dephasing} shows the resulting dephasing for our fiducial $10+10\Msun$ binary, expressed as the number of cycles lost relative to a vacuum inspiral, $\delta N(>\!f)=N_{\rm vac}-N_{\rm star}$, as a function of the reference frequency $f$. The magnitude of the effect is large, but its observability depends strongly on the band. In the LISA band the gas drag is so strong that the binary never completes a quasi-stationary inspiral: it is swept through the milli-Hz regime toward merger, so the nominal $\delta N\simeq N_{\rm V}$ there reflects a rapid, gas-driven sweep through the band---a ``plunge'' in this sense, not a relativistic one---rather than a slowly-accumulating, measurable phase offset, consistent with the negligible characteristic strain estimated above. Near the LVK band, conversely, gas drag has fallen well below GW emission and the binary merges almost as in vacuum, leaving only a modest residual after the coalescence time and phase are removed---of order ten cycles above $10$~Hz for a gas-only sound speed, and a few cycles for the radiation-dominated adiabatic sound speed of the envelope. This is a raw phase residual: marginalizing over the binary masses and spins can absorb part of it, so we treat it as an upper bound on the cleanly measurable dephasing. The largest measurable departure from a vacuum waveform therefore falls in the intervening deci-Hz band, where the source is simultaneously loud and still gas-dominated.

This is made explicit by the characteristic-strain track of the source across the detector noise curves (Figure~\ref{fig:hstrain}). Gas hardening suppresses $h_{\rm c}$ by several orders of magnitude in the LISA band, because the inspiral cycles that would otherwise accumulate there are erased by the rapid sweep. The track then recovers toward its vacuum value as the binary crosses the deci-Hz and LVK bands on its way to merger. An immortal-channel merger is thus not a persistent milli-Hz LISA source but a deci-Hz transient that ends as an ordinary-looking LVK merger; a $10+10\Msun$ merger at $100$~Mpc accumulates a sky-averaged LVK signal-to-noise of $\sim50$, comparable to a typical detected event. In the deci-Hz band the source is far louder still: the same gas-hardened binary at $100$~Mpc retains a sky-averaged DECIGO signal-to-noise of $\sim1.2\times10^{3}$, an order of magnitude below the $\sim1.1\times10^{4}$ of the corresponding vacuum inspiral. The gas imprint at deci-Hz is therefore not a subtle phase residual but an order-of-magnitude suppression of an extremely loud signal. The detailed waveform in this band inherits the nonlinear-core uncertainties of \S\ref{sec:bbh}, so the magnitude of the imprint is robust while its precise shape is not. Resolving it requires sensitivity in the deci-Hz decade, motivating next-generation concepts such as DECIGO \citep{Kawamura11}.

\subsection{Observables: EM}
Immortal stars that migrate directly onto the SMBH could yield Roche-lobe overflow onto the SMBH \citep{2022ApJ...929..133J} or, if perturbed onto sufficiently eccentric orbits, tidal disruption events (TDEs) around high-mass SMBHs, which might account for the sudden flares observed in compact radio galactic nuclei \citep{Readhead24}. (A BBH merger kick imparts only $\sim(M_{\rm BBH}/M_{\ast})\,v_{\rm kick}\lesssim{\rm tens~of~km/s}$ of recoil to the star itself, a percent-level orbital perturbation, so delivery to the SMBH must rely on migration or dynamical scattering rather than merger recoil.)

The merger itself should also have an electromagnetic counterpart. It occurs inside, or within hours of entering, a dense stellar interior: the hardening binary and the kicked remnant accrete at hyper-Eddington, photon-trapped rates with the energy escaping mechanically in jets and winds (\S\ref{sec:orbital}), and the hardening deposits of order the star's net binding energy (\S\ref{sec:noimmortal}). The natural expectation is therefore an AGN flare coincident with an LVK BBH merger, or promptly following it after the jet/envelope breakout time, analogous to the flare candidates proposed as AGN-disk counterparts to LVK events \citep{Graham20}. A quantitative lightcurve prediction requires the radiation-hydrodynamic modeling flagged above and is left to future work.

More dramatically, the destruction of an embedded massive star is a natural progenitor for the most luminous nuclear transients, whether a single compact object drives it to a runaway, hyper-Eddington accretion event (\S\ref{sec:orbital}) or the star is tidally disrupted as it is perturbed toward the SMBH. The recently reported flare of \citet{2026NatAs..10..154G}, which radiated $\sim10^{54}\,$erg and is most plausibly the shredding of a $\gtrsim30\,\Msun$ star, sits at the energy scale expected when a substantial fraction of a massive star is consumed in such an event; immortal stars ($M_{\ast}\sim10^{2}$--$10^{3}\,\Msun$) provide a ready supply of the massive progenitors that such flares require. The in-plane tidal disruption of stars within AGN disks has been simulated in detail \citep{2024MNRAS.527.8103R}, and the capture and infall of stars that supply both the embedded compact objects and the disruptable stellar population were examined by \citet{2022MNRAS.514.4102M}.

\section{Conclusions}\label{sec:conclusion}
Immortal stars in AGN disks act as binary black hole formation accelerants. Gas drag within immortal stars rapidly drives low-mass BHs together to form BBHs that are hard to most dynamical encounters in the disk and biased to prograde orientations. Even the fiducial $10+10\,\Msun$ merger deposits of order the star's net binding energy during its final hardening, so each immortal likely survives only one or two mergers before being strongly heated or unbound, redistributing its mass into the AGN disk midplane in turbulent, mixing events. The channel therefore matters mostly for first-generation, lower-mass BH mergers. A single compact object spiraling into an immortal leads either to a collapsar-like, engine-driven transient, if accretion proceeds near the hyper-Eddington BHL rate, or to a long-lived, quasi-star-like configuration, if mechanical feedback regulates the inflow.

The BBH mergers themselves are loud LVK sources, but their distinctive signature---the suppression and dephasing of the inspiral by the dense gas---is a deci-Hz phenomenon, best probed by future detectors such as DECIGO. In favorable scenarios the channel could contribute up to $\sim8\,{\rm Gpc^{-3}\,yr^{-1}}$ to the BBH merger rate, though this ceiling carries order-of-magnitude downward uncertainty (\S\ref{sec:rates}). The central modeling uncertainty, the strongly nonlinear coupling between the hardening binary and the surrounding gas, can ultimately be resolved only with 3D self-gravitating simulations.
\section*{Software}
\texttt{MESA} \citep[][\url{https://docs.mesastar.org}]{{2011ApJS..192....3P},{2013ApJS..208....4P},{2015ApJS..220...15P},{2018ApJS..234...34P},{2019ApJS..243...10P},{2022arXiv220803651J}},
\texttt{MESASDK} \citep{mesasdk_linux},
\texttt{matplotlib} \citep{4160265}, \texttt{numpy} \citep{5725236}, and the \texttt{agn\_cee} reproducibility package (\url{https://github.com/matteocantiello/agnstars_cee}), which contains the code, input data, and notebooks reproducing every dynamical calculation and figure in \S\ref{sec:results}--\ref{sec:discussion}.

\section*{Acknowledgments}
The Center for Computational Astrophysics at the Flatiron Institute is supported by the Simons Foundation.
KESF \& BM are supported by NSF AST-2206096, NSF AST-1831415 and Simons Foundation Grant 533845.
A.J.D. was supported by NASA through the Hubble Fellowship Program grant No.~HST-HF2-51553.001, awarded by the Space Telescope Science Institute, which is operated by the Association of Universities for Research in Astronomy, Inc., for NASA, under contract NAS5-26555.
\appendix

\section{Software details}\label{apndx:A}
Calculations were carried out using \texttt{MESA} version r22.05.1.

The \texttt{MESA} EOS is a blend of the OPAL \citep{Rogers2002}, SCVH
\citep{Saumon1995}, FreeEOS \citep{Irwin2004}, HELM \citep{Timmes2000},
PC \citep{Potekhin2010}, and Skye \citep{2021ApJ...913...72J} EOSes.

Radiative opacities are primarily from OPAL \citep{Iglesias1993,
Iglesias1996}, with low-temperature data from \citet{Ferguson2005}
and the high-temperature, Compton-scattering dominated regime by
\citet{Poutanen2017}.  Electron conduction opacities are from
\citet{Cassisi2007}.

Nuclear reaction rates are from JINA REACLIB \citep{Cyburt2010}, NACRE \citep{Angulo1999} and
additional tabulated weak reaction rates \citep{Fuller1985, Oda1994,
Langanke2000}.  Screening is included via the prescription of \citet{Chugunov2007}.
Thermal neutrino loss rates are from \citet{Itoh1996}.

We adopted a 21-isotope nuclear network (approx21.net).
We used the Schwarzschild criterion to determine convective
boundaries and did not include convective overshooting.

\section{Gas-hardening model ladder}\label{apndx:ladder}
\begin{figure}
\centering
\includegraphics[width=\columnwidth]{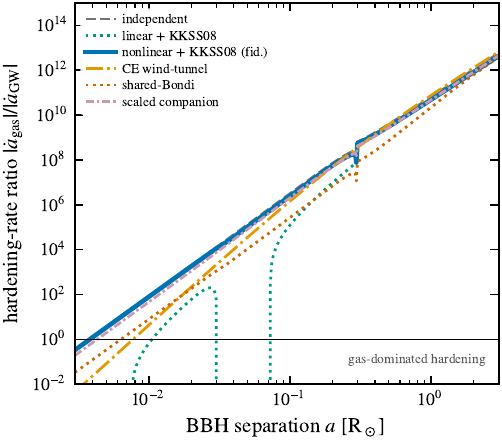}
\caption{Ladder of gas-hardening prescriptions for the internal BBH orbit: the ratio of the gas-driven to the GW-driven hardening rate, $|\dot a_{\rm gas}|/|\dot a_{\rm GW}|$, as a function of internal separation, for a $10+10\Msun$ binary at the stellar core. The fiducial model (nonlinear self-wake $+$ the \citet{KKSS08} companion wake) is the heavy curve. For the physically motivated prescriptions, gas hardening exceeds GW emission by many orders of magnitude over the separations that dominate the residence time; the spread between rungs sets the dominant modeling uncertainty, and only the pure-linear rung drops toward the gas--GW parity line (the cautionary near-stalling limit, excluded from the quoted band).}
\label{fig:ladder}
\end{figure}

Our central claim, that gas drag rather than GW emission governs the late hardening of the internal BBH (\S\ref{sec:bbh}), rests on a ladder of dynamical-friction prescriptions whose spread we treat as the dominant modeling uncertainty. Figure~\ref{fig:ladder} compares them directly, showing the ratio of the gas-driven to the GW-driven hardening rate, $|\dot a_{\rm gas}|/|\dot a_{\rm GW}|$, as a function of internal separation for all six rungs. For all physically motivated prescriptions, gas hardening exceeds GW emission by many orders of magnitude over the separations that dominate the residence time, so the qualitative conclusion is insensitive to the choice of model; only near the final GW-dominated regime, or in the deliberately cautionary pure-linear rung, does this cease to hold. The rungs differ mainly in normalization, with the pure-linear prescription dropping toward the gas--GW parity line at small separation. (In that rung the weak supersonic self-wake is over-cancelled by the companion term; we floor it at stalling rather than permit gas-driven anti-hardening, and exclude it from the quoted band.)

The six rungs are summarized in Table~\ref{tab:ladder}. Each treats the per-component drag force $F_{i}$ acting at the orbital speed $v_{i}=v_{\rm rel}\,m_{j}/M$ and radius $r_{i}=a\,m_{j}/M$ about the center of mass, with the gas energy-loss rate $L_{\rm gas}=\sum_{i}(F_{i}+F_{{\rm acc},i})v_{i}$. The companion-wake contribution is taken from the verbatim \citet{KKSS08} azimuthal drag coefficient, which is negative---the trailing companion wake exerts a forward torque that partially cancels the backward self-wake drag:
\begin{equation}
I_{2,\phi}(\mathcal{M}) = \mathcal{M}^{2}\times
\begin{cases}
-0.022\,(10-\mathcal{M})\tanh(1.5\,\mathcal{M}), & \mathcal{M}<2.97,\\[3pt]
-0.13 + 0.07\arctan(5\mathcal{M}-15), & \mathcal{M}\geq2.97,
\end{cases}
\label{eq:kkss08}
\end{equation}
entering the force as the additive (sign-preserved) term $F_{\rm comp}=4\pi\rho\,(Gm_{i})^{2}\,I_{2,\phi}(\mathcal{M})/v_{i}^{2}$.

Two rungs use parametrized factors not defined elsewhere. Model 4 multiplies the fiducial rate by a shared-Bondi suppression factor $f_{\rm sB}(\chi_{\rm B}) = 0.1 + 0.9/(1+\chi_{\rm B}^{2})$ with $\chi_{\rm B}=R_{\rm B,BBH}/a$, interpolating from no suppression ($\chi_{\rm B}\ll1$) to a floor of $0.1$ deep in the shared-Bondi regime. Model 5 scales the nonlinear self-wake by $(1-\eta_{\rm comp})$ with $\eta_{\rm comp}\simeq0.45$, the supersonic linear companion/self cancellation fraction. We note that the fiducial Model 2 grafts the linear \citet{KKSS08} companion term onto the nonlinear self-wake in a regime ($R_{\rm BHL,i}/r_{i}\gtrsim1$) where linear double-wake theory does not strictly apply. Its impact is small, however: Models 0 (no companion term) and 5 bracket it, and both differ from the fiducial merger time by $\lesssim40\%$. Linear-response calculations in stratified envelopes \citep{Gagnier26} show that the companion wake has no local Coulomb-logarithmic term. As the self-wake is resolved down to smaller scales its drag keeps growing while the companion term does not, so the cancellation becomes fractionally smaller. This further supports excluding the over-cancelled pure-linear rung.

\begin{table*}
\centering
\caption{Ladder of gas-hardening prescriptions for the internal BBH orbit. Merger times are integrated from $a_{\rm BBH}=R_{\odot}$ for an equal-mass $10+10\,\Msun$ binary at the stellar core; the vacuum (GW-only) merger time from the same separation is $7.5\times10^{4}$~yr. The quoted band (Models 0, 2, 3, 4) spans $\sim1$--$8$~hr.}
\label{tab:ladder}
\begin{tabular}{cllcc}
\hline\hline
Model & Prescription & Drag law / coefficient & In band? & $t_{\rm merge}$ \\
\hline
0 & independent wake / upper-drag baseline & nonlinear self-wake \citep{Kim10}, no companion & yes & $1.0$~hr \\
1 & linear self $+$ KKSS08 companion & linear self-wake \citep{Kim207} $+$ Eq.~(\ref{eq:kkss08}) & no$^{a}$ & $2$~yr \\
2 & nonlinear self $+$ KKSS08 (fiducial) & nonlinear self-wake \citep{Kim10} $+$ Eq.~(\ref{eq:kkss08}) & yes & $1.1$~hr \\
3 & CE wind-tunnel $C_{d}$ & $F=C_{d}(\mathcal{M})\,\pi R_{\rm a}^{2}\rho v^{2}$ (wind-tunnel) & yes & $6.7$~hr \\
4 & shared-Bondi suppressed & Model 2 $\times\,f_{\rm sB}(R_{\rm B,BBH}/a)$ & yes & $7.5$~hr \\
5 & scaled companion (stress test) & nonlinear self $\times\,(1-\eta_{\rm comp})$ & no$^{b}$ & $1.5$~hr \\
\hline
\end{tabular}\\[3pt]
\begin{minipage}{0.92\textwidth}
\footnotesize $^{a}$The pure-linear rung is not treated as a physical prediction in the core: it is shown only to illustrate the breakdown of linear double-wake theory when $R_{\rm BHL,i}/r_{i}\gtrsim1$, where the exact companion term over-cancels the weak linear self-wake. $^{b}$Model 5 is a stress-test extrapolation---it assumes the linear companion/self cancellation ratio survives into the nonlinear regime---and is excluded from the quoted band. Models 3 and 4 are deliberate systematic brackets: common-envelope wind-tunnel drag coefficients \citep{MacLeod17,De20} and adiabatic shared-Bondi suppression, respectively.
\end{minipage}
\end{table*}

\clearpage
\bibliographystyle{aasjournal}
\bibliography{references}

\end{document}